\input harvmac
\input epsf

\def\hat{\widehat}

%
\let\includefigures=\iftrue
%
%
%
\newfam\black
\input rotate
\input epsf
\noblackbox
%
%
\includefigures
\message{If you do not have epsf.tex (to include figures),}
\message{change the option at the top of the tex file.}
\def\figin{\epsfcheck\figin}\def\figins{\epsfcheck\figins}
\def\epsfcheck{\ifx\epsfbox\UnDeFiNeD
\message{(NO epsf.tex, FIGURES WILL BE IGNORED)}
\gdef\figin##1{\vskip2in}\gdef\figins##1{\hskip.5in}
\else\message{(FIGURES WILL BE INCLUDED)}%
\gdef\figin##1{##1}\gdef\figins##1{##1}\fi}
\def\DefWarn#1{}
\def\N{{\cal N}}
\def\figinsert{\goodbreak\midinsert}
\def\ifig#1#2#3{\DefWarn#1\xdef#1{fig.~\the\figno}
\writedef{#1\leftbracket fig.\noexpand~\the\figno}%
\figinsert\figin{\centerline{#3}}\medskip\centerline{\vbox{\baselineskip12pt
\advance\hsize by -1truein\noindent\footnotefont{\bf
Fig.~\the\figno:} #2}}
\bigskip\endinsert\global\advance\figno by1}
\else
\def\ifig#1#2#3{\xdef#1{fig.~\the\figno}
\writedef{#1\leftbracket fig.\noexpand~\the\figno}%
\global\advance\figno by1} \fi
\def\hat{\widehat}
\def\tilde{\widetilde}

\def\yboxit#1#2{\vbox{\hrule height #1 \hbox{\vrule width #1
\vbox{#2}\vrule width #1 }\hrule height #1 }}
\def\fillbox#1{\hbox to #1{\vbox to #1{\vfil}\hfil}}
\def\ybox{{\lower 1.3pt \yboxit{0.4pt}{\fillbox{8pt}}\hskip-0.2pt}}

\def\rightarrowbox#1#2{
  \setbox1=\hbox{\kern#1{${ #2}$}\kern#1}
  \,\vbox{\offinterlineskip\hbox to\wd1{\hfil\copy1\hfil}
    \kern 3pt\hbox to\wd1{\rightarrowfill}}}

\def\half{{1\over 2}}

\def\vev#1{\langle{#1}\rangle}

\def\tilde{\widetilde}

\def\II{\relax{I\kern-.10em I}}

\def\pt{pt}
\def\bar{\overline}

\def\IZ{\relax\ifmmode\mathchoice
{\hbox{\cmss Z\kern-.4em Z}}{\hbox{\cmss Z\kern-.4em Z}}
{\lower.9pt\hbox{\cmsss Z\kern-.4em Z}} {\lower1.2pt\hbox{\cmsss
Z\kern-.4em Z}}\else{\cmss Z\kern-.4em Z}\fi}
\def\IB{\relax{\rm I\kern-.18em B}}
\def\IC{{\relax\hbox{$\inbar\kern-.3em{\rm C}$}}}
\def\ID{\relax{\rm I\kern-.18em D}}
\def\IE{\relax{\rm I\kern-.18em E}}
\def\IF{\relax{\rm I\kern-.18em F}}
\def\IG{\relax\hbox{$\inbar\kern-.3em{\rm G}$}}
\def\IGa{\relax\hbox{${\rm I}\kern-.18em\Gamma$}}
\def\IH{\relax{\rm I\kern-.18em H}}
\def\II{\relax{\rm I\kern-.18em I}}
\def\IK{\relax{\rm I\kern-.18em K}}
\def\IN{\relax{\rm I\kern-.18em N}}
\def\IP{\relax{\rm I\kern-.18em P}}

%
\def\inbar{\,\vrule height1.5ex width.4pt depth0pt}

\font\cmss=cmss10 \font\cmsss=cmss10 at 7pt
\def\IR{\relax{\rm I\kern-.18em R}}

\def\lp10{l_P^{10}}
\def\lp11{l_P^{11}}
\def\R11{R_{11}}

\def\a{\alpha}
\def\b{\beta}
\def\lt{\tilde\lambda}

\def\gb#1{ {\langle #1 ] } }

\newbox\tmpbox\setbox\tmpbox\hbox{\abstractfont
}
 \Title{\vbox{\baselineskip12pt\hbox to\wd\tmpbox{\hss
 hep-th/0412308} }}
 {\vbox{\centerline{New Recursion Relations }
 \bigskip
 \centerline{for Tree Amplitudes of Gluons}
 }}
\smallskip
\centerline{Ruth Britto, Freddy Cachazo and Bo Feng}
\smallskip
\bigskip
\centerline{\it School of Natural Sciences, Institute for Advanced
Study, Princeton NJ 08540 USA}
\bigskip
\vskip 1cm \noindent

\input amssym.tex

We present new recursion relations for tree amplitudes in gauge
theory that give very compact formulas. Our relations give any
tree amplitude as a sum over terms constructed from products of
two amplitudes of fewer particles multiplied by a Feynman
propagator.
 The two amplitudes in each term are
physical, in the sense that all particles are on-shell and
momentum conservation is preserved.
 This is striking, since it is just like
adding certain factorization limits of the original amplitude to
build up the full answer. As examples, we recompute all known
tree-level amplitudes of up to seven gluons and show that our
recursion relations naturally give their most compact forms. We
give a new result for an eight-gluon amplitude,
$A(1^+,2^-,3^+,4^-,5^+,6^-,7^+,8^-)$.  We
show how to build any amplitude in terms of three-gluon amplitudes only.

\Date{December 2004}

\lref\BernZX{ Z.~Bern, L.~J.~Dixon, D.~C.~Dunbar and
D.~A.~Kosower, ``One Loop N Point Gauge Theory Amplitudes,
Unitarity And Collinear Limits,'' Nucl.\ Phys.\ B {\bf 425}, 217
(1994), hep-ph/9403226.
}

\lref\BernCG{ Z.~Bern, L.~J.~Dixon, D.~C.~Dunbar and
D.~A.~Kosower, ``Fusing Gauge Theory Tree Amplitudes into Loop
Amplitudes,'' Nucl.\ Phys.\ B {\bf 435}, 59 (1995),
hep-ph/9409265.
}

\lref\WittenNN{
E.~Witten,
``Perturbative gauge theory as a string theory in twistor space,''
Commun.\ Math.\ Phys.\  {\bf 252}, 189 (2004)
[arXiv:hep-th/0312171].
}

\lref\CachazoKJ{
F.~Cachazo, P.~Svrcek and E.~Witten,
``MHV vertices and tree amplitudes in gauge theory,''
JHEP {\bf 0409}, 006 (2004)
[arXiv:hep-th/0403047].
}

\lref\BerkovitsJJ{
N.~Berkovits and E.~Witten,
``Conformal supergravity in twistor-string theory,''
JHEP {\bf 0408}, 009 (2004)
[arXiv:hep-th/0406051].
}

\lref\penrose{R. Penrose, ``Twistor Algebra,'' J. Math. Phys. {\bf
8} (1967) 345.}

\lref\berends{F. A. Berends, W. T. Giele and H. Kuijf, ``On
Relations Between Multi-Gluon And Multi-Graviton Scattering,"
Phys. Lett {\bf B211} (1988) 91.}

\lref\berendsgluon{F. A. Berends, W. T. Giele and H. Kuijf,
``Exact and Approximate Expressions for Multigluon Scattering,"
Nucl. Phys. {\bf B333} (1990) 120.}

\lref\bernplusa{Z. Bern, L. Dixon and D. A. Kosower, ``New QCD
Results From String Theory,'' in {\it Strings '93}, ed. M. B.
Halpern et. al. (World-Scientific, 1995), hep-th/9311026.}

\lref\bernplusb{Z. Bern, G. Chalmers, L. J. Dixon and D. A.
Kosower, ``One Loop $N$ Gluon Amplitudes with Maximal Helicity
Violation via Collinear Limits," Phys. Rev. Lett. {\bf 72} (1994)
2134.}

\lref\bernfive{Z. Bern, L. J. Dixon and D. A. Kosower, ``One Loop
Corrections to Five Gluon Amplitudes," Phys. Rev. Lett {\bf 70}
(1993) 2677.}

\lref\bernfourqcd{Z.Bern and  D. A. Kosower, "The Computation of
Loop Amplitudes in Gauge Theories," Nucl. Phys.  {\bf B379,}
(1992) 451.}

\lref\cremmerlag{E. Cremmer and B. Julia, ``The $N=8$ Supergravity
Theory. I. The Lagrangian," Phys. Lett.  {\bf B80} (1980) 48.}

\lref\cremmerso{E. Cremmer and B. Julia, ``The $SO(8)$
Supergravity," Nucl. Phys.  {\bf B159} (1979) 141.}

\lref\dewitt{B. DeWitt, "Quantum Theory of Gravity, III:
Applications of Covariant Theory," Phys. Rev. {\bf 162} (1967)
1239.}

\lref\dunbarn{D. C. Dunbar and P. S. Norridge, "Calculation of
Graviton Scattering Amplitudes Using String Based Methods," Nucl.
Phys. B {\bf 433,} 181 (1995), hep-th/9408014.}

\lref\ellissexton{R. K. Ellis and J. C. Sexton, "QCD Radiative
corrections to parton-parton scattering," Nucl. Phys.  {\bf B269}
(1986) 445.}

\lref\gravityloops{Z. Bern, L. Dixon, M. Perelstein, and J. S.
Rozowsky, ``Multi-Leg One-Loop Gravity Amplitudes from Gauge
Theory,"  hep-th/9811140.}

\lref\kunsztqcd{Z. Kunszt, A. Singer and Z. Tr\'{o}cs\'{a}nyi,
``One-loop Helicity Amplitudes For All $2\rightarrow2$ Processes
in QCD and ${\cal N}=1$ Supersymmetric Yang-Mills Theory,'' Nucl.
Phys.  {\bf B411} (1994) 397, hep-th/9305239.}

\lref\mahlona{G. Mahlon, ``One Loop Multi-photon Helicity
Amplitudes,'' Phys. Rev.  {\bf D49} (1994) 2197, hep-th/9311213.}

\lref\mahlonb{G. Mahlon, ``Multi-gluon Helicity Amplitudes
Involving a Quark Loop,''  Phys. Rev.  {\bf D49} (1994) 4438,
hep-th/9312276.}

\lref\klt{H. Kawai, D. C. Lewellen and S.-H. H. Tye, ``A Relation
Between Tree Amplitudes of Closed and Open Strings," Nucl. Phys.
{B269} (1986) 1.}

\lref\pppmgr{Z. Bern, D. C. Dunbar and T. Shimada, ``String Based
Methods In Perturbative Gravity," Phys. Lett.  {\bf B312} (1993)
277, hep-th/9307001.}

\lref\GiombiIX{ S.~Giombi, R.~Ricci, D.~Robles-Llana and
D.~Trancanelli, ``A Note on Twistor Gravity Amplitudes,''
hep-th/0405086.
}

\lref\WuFB{ J.~B.~Wu and C.~J.~Zhu, ``MHV Vertices and Scattering
Amplitudes in Gauge Theory,'' hep-th/0406085.
}

\lref\Feynman{R.P. Feynman, Acta Phys. Pol. 24 (1963) 697, and in
{\it Magic Without Magic}, ed. J. R. Klauder (Freeman, New York,
1972), p. 355.}

\lref\Peskin{M.~E. Peskin and D.~V. Schroeder, {\it An Introduction
to Quantum Field Theory} (Addison-Wesley Pub. Co., 1995).}

\lref\parke{S. Parke and T. Taylor, ``An Amplitude For $N$ Gluon
Scattering,'' Phys. Rev. Lett. {\bf 56} (1986) 2459. }

\lref\berandgiele{F. A. Berends
and W. T. Giele, ``Recursive Calculations For Processes With $N$
Gluons,'' Nucl. Phys. {\bf B306} (1988) 759.}

\lref\BrandhuberYW{ A.~Brandhuber, B.~Spence and G.~Travaglini,
``One-Loop Gauge Theory Amplitudes In N = 4 Super Yang-Mills From
MHV Vertices,'' hep-th/0407214.
}

\lref\CachazoZB{ F.~Cachazo, P.~Svr\v cek and E.~Witten, ``Twistor
space structure of one-loop amplitudes in gauge theory,''
hep-th/0406177.
}

\lref\passarino{ L.~M. Brown and R.~P. Feynman, ``Radiative Corrections To Compton Scattering,'' Phys. Rev. 85:231
(1952); G.~Passarino and M.~Veltman, ``One Loop Corrections For E+ E- Annihilation Into Mu+ Mu- In The Weinberg
Model,'' Nucl. Phys. B160:151 (1979);
G.~'t Hooft and M.~Veltman, ``Scalar One Loop Integrals,'' Nucl. Phys. B153:365 (1979); R.~G.~
Stuart, ``Algebraic Reduction Of One Loop Feynman Diagrams To Scalar Integrals,'' Comp. Phys. Comm. 48:367 (1988); R.~G.~Stuart and A.~Gongora, ``Algebraic Reduction Of One Loop Feynman Diagrams To Scalar Integrals. 2,'' Comp. Phys. Comm. 56:337 (1990).}

\lref\neerven{ W. van Neerven and J. A. M. Vermaseren, ``Large Loop Integrals,'' Phys. Lett.
137B:241 (1984)}

\lref\melrose{ D.~B.~Melrose, ``Reduction Of Feynman Diagrams,'' Il Nuovo Cimento 40A:181 (1965); G.~J.~van Oldenborgh and J.~A.~M.~Vermaseren, ``New Algorithms For One Loop Integrals,'' Z. Phys. C46:425 (1990);
G.J. van Oldenborgh,  PhD Thesis, University of Amsterdam (1990);
A. Aeppli, PhD thesis, University of Zurich (1992).}

\lref\bernTasi{Z.~Bern, hep-ph/9304249, in {\it Proceedings of
Theoretical Advanced Study Institute in High Energy Physics (TASI
92)}, eds. J. Harvey and J. Polchinski (World Scientific, 1993). }

\lref\morgan{ Z.~Bern and A.~G.~Morgan, ``Supersymmetry relations
between contributions to one loop gauge boson amplitudes,'' Phys.\
Rev.\ D {\bf 49}, 6155 (1994), hep-ph/9312218.
}

\lref\RoiSpV{R.~Roiban, M.~Spradlin and A.~Volovich, ``A Googly
Amplitude From The B-Model In Twistor Space,'' JHEP {\bf 0404},
012 (2004) hep-th/0402016; R.~Roiban and A.~Volovich, ``All Googly
Amplitudes From The $B$-Model In Twistor Space,'' hep-th/0402121;
R.~Roiban, M.~Spradlin and A.~Volovich, ``On The Tree-Level
S-Matrix Of Yang-Mills Theory,'' Phys.\ Rev.\ D {\bf 70}, 026009
(2004) hep-th/0403190}

\lref\CachazoBY{ F.~Cachazo, P.~Svr\v cek and E.~Witten, ``Gauge
Theory Amplitudes In Twistor Space And Holomorphic Anomaly,''
hep-th/0409245.
}

\lref\DixonWI{ L.~J.~Dixon, ``Calculating Scattering Amplitudes
Efficiently,'' hep-ph/9601359.
}

\lref\BernMQ{ Z.~Bern, L.~J.~Dixon and D.~A.~Kosower, ``One Loop
Corrections To Five Gluon Amplitudes,'' Phys.\ Rev.\ Lett.\  {\bf
70}, 2677 (1993), hep-ph/9302280.
}

\lref\berends{F.~A.~Berends, R.~Kleiss, P.~De Causmaecker, R.~Gastmans and T.~T.~Wu, ``Single Bremsstrahlung Processes In Gauge Theories,'' Phys. Lett. {\bf B103} (1981) 124; P.~De
Causmaeker, R.~Gastmans, W.~Troost and T.~T.~Wu, ``Multiple Bremsstrahlung In Gauge Theories At High-Energies. 1. General
Formalism For Quantum Electrodynamics,'' Nucl. Phys. {\bf
B206} (1982) 53; R.~Kleiss and W.~J.~Stirling, ``Spinor Techniques For Calculating P Anti-P $\to$ W+- / Z0 + Jets,'' Nucl. Phys. {\bf
B262} (1985) 235; R.~Gastmans and T.~T. Wu, {\it The Ubiquitous
Photon: Heliclity Method For QED And QCD} Clarendon Press, 1990.}

\lref\xu{Z. Xu, D.-H. Zhang and L. Chang, ``Helicity Amplitudes For Multiple
Bremsstrahlung In Massless Nonabelian Theories,''
 Nucl. Phys. {\bf B291}
(1987) 392.}

\lref\gunion{J.~F. Gunion and Z. Kunszt, ``Improved Analytic Techniques For Tree Graph Calculations And The G G Q
Anti-Q Lepton Anti-Lepton Subprocess,''
Phys. Lett. {\bf 161B}
(1985) 333.}

\lref\GeorgiouBY{ G.~Georgiou, E.~W.~N.~Glover and V.~V.~Khoze,
``Non-MHV Tree Amplitudes In Gauge Theory,'' JHEP {\bf 0407}, 048
(2004), hep-th/0407027.
}

\lref\WuJX{ J.~B.~Wu and C.~J.~Zhu, ``MHV Vertices And Fermionic
Scattering Amplitudes In Gauge Theory With Quarks And Gluinos,''
hep-th/0406146.
}

\lref\WuFB{ J.~B.~Wu and C.~J.~Zhu, ``MHV Vertices And Scattering
Amplitudes In Gauge Theory,'' JHEP {\bf 0407}, 032 (2004),
hep-th/0406085.
}

\lref\GeorgiouWU{ G.~Georgiou and V.~V.~Khoze, ``Tree Amplitudes
In Gauge Theory As Scalar MHV Diagrams,'' JHEP {\bf 0405}, 070
(2004), hep-th/0404072.
}

\lref\Nair{V. Nair, ``A Current Algebra For Some Gauge Theory
Amplitudes," Phys. Lett. {\bf B78} (1978) 464. }

\lref\BernAD{ Z.~Bern, ``String Based Perturbative Methods For
Gauge Theories,'' hep-ph/9304249.
}

\lref\BernKR{ Z.~Bern, L.~J.~Dixon and D.~A.~Kosower,
``Dimensionally Regulated Pentagon Integrals,'' Nucl.\ Phys.\ B
{\bf 412}, 751 (1994), hep-ph/9306240.
}

\lref\CachazoDR{ F.~Cachazo, ``Holomorphic Anomaly Of Unitarity
Cuts And One-Loop Gauge Theory Amplitudes,'' hep-th/0410077.
}

\lref\irref{W. T. Giele and E. W. N. Glover, ``Higher order corrections to jet cross-sections in e+ e- annihilation,'' Phys. Rev. {\bf D46}
(1992) 1980; Z. Kunszt, A. Signer and Z. Tr\' ocs\' anyi, ``Singular terms of helicity amplitudes at one loop in QCD and the soft limit
of the cross-sections of multiparton processes,'' Nucl. Phys. {\bf
B420} (1994) 550.}

\lref\giel{W. T. Giele and E. W. N. Glover, ``Higher order corrections to jet cross-sections in e+ e- annihilation,'' Phys. Rev. {\bf D46}
(1992) 1980; W. T. Giele, E. W. N. Glover and D. A. Kosower, ``Higher order corrections to jet cross-sections in hadron colliders,'' Nucl.
Phys. {\bf B403} (1993) 633. }

\lref\kuni{Z. Kunszt and D. Soper, ``Calculation of jet cross-sections in hadron collisions at order alpha-s**3,''Phys. Rev. {\bf D46} (1992)
192; Z. Kunszt, A. Signer and Z. Tr\' ocs\' anyi, ``Singular terms of helicity amplitudes at one loop in QCD and the soft limit
of the cross-sections of multiparton processes,'' Nucl. Phys. {\bf
B420} (1994) 550. }

\lref\seventree{F.~A. Berends, W.~T. Giele and H. Kuijf, ``Exact And Approximate Expressions For Multi - Gluon Scattering,'' Nucl. Phys.
{\bf B333} (1990) 120.}

\lref\mangpxu{M. Mangano, S.~J. Parke and Z. Xu, ``Duality And Multi - Gluon Scattering,'' Nucl. Phys. {\bf B298}
(1988) 653.}

\lref\mangparke{M. Mangano and S.~J. Parke, ``Multiparton Amplitudes In Gauge Theories,'' Phys. Rep. {\bf 200}
(1991) 301.}

\lref\grisaru{M. T. Grisaru, H. N. Pendleton and P. van Nieuwenhuizen, ``Supergravity And The S Matrix,'' Phys. Rev.  {\bf D15} (1977) 996; M. T. Grisaru and H. N. Pendleton, ``Some Properties Of Scattering Amplitudes In Supersymmetric Theories,'' Nucl. Phys. {\bf B124} (1977) 81.}

\lref\Bena{I. Bena, Z. Bern, D. A. Kosower and R. Roiban, ``Loops in Twistor Space,'' hep-th/0410054.}

\lref\BernKY{
Z.~Bern, V.~Del Duca, L.~J.~Dixon and D.~A.~Kosower,
``All Non-Maximally-Helicity-Violating One-Loop Seven-Gluon Amplitudes In N =
4 Super-Yang-Mills Theory,''
arXiv:hep-th/0410224.
}

\lref\BrittoNJ{
R.~Britto, F.~Cachazo and B.~Feng,
``Computing one-loop amplitudes from the holomorphic anomaly of unitarity
cuts,''
arXiv:hep-th/0410179.
}

\lref\BidderTX{
S.~J.~Bidder, N.~E.~J.~Bjerrum-Bohr, L.~J.~Dixon and D.~C.~Dunbar,
``N = 1 supersymmetric one-loop amplitudes and the holomorphic anomaly of
unitarity cuts,''
arXiv:hep-th/0410296.
}

\lref\DennerQQ{
A.~Denner, U.~Nierste and R.~Scharf,
``A Compact expression for the scalar one loop four point function,''
Nucl.\ Phys.\ B {\bf 367}, 637 (1991).
}

\lref\BrittoTX{
R.~Britto, F.~Cachazo and B.~Feng,
``Coplanarity in twistor space of N = 4 next-to-MHV one-loop amplitude
coefficients,''
arXiv:hep-th/0411107.
}

\lref\BidderVX{
S.~J.~Bidder, N.~E.~J.~Bjerrum-Bohr, D.~C.~Dunbar and W.~B.~Perkins,
``Twistor Space Structure of the Box Coefficients of N=1 One-loop
Amplitudes,''
arXiv:hep-th/0412023.
}

\lref\BrittoNJ{
R.~Britto, F.~Cachazo and B.~Feng,
``Computing one-loop amplitudes from the holomorphic anomaly of unitarity
cuts,''
arXiv:hep-th/0410179.
}
\lref\KosowerYZ{
D.~A.~Kosower,
``Next-to-maximal helicity violating amplitudes in gauge theory,''
arXiv:hep-th/0406175.
}

\lref\BernKY{
Z.~Bern, V.~Del Duca, L.~J.~Dixon and D.~A.~Kosower,
``All non-maximally-helicity-violating one-loop seven-gluon amplitudes in
N =
4 super-Yang-Mills theory,''
arXiv:hep-th/0410224.
}
                                                                                \lref\sbook{
R. J. Eden, P. V. Landshoff, D. I. Olive and J. C. Polkinghorne,
{\it The Analytic S-Matrix}, Cambridge University Press, 1966.
}

\lref\BernSC{
Z.~Bern, L.~J.~Dixon and D.~A.~Kosower,
``One-loop amplitudes for e+ e- to four partons,''
Nucl.\ Phys.\ B {\bf 513}, 3 (1998)
[arXiv:hep-ph/9708239].
}

\lref\mhvdiag{
C.~J.~Zhu,
``The googly amplitudes in gauge theory,''
JHEP {\bf 0404}, 032 (2004)
[arXiv:hep-th/0403115].
G.~Georgiou and V.~V.~Khoze,
``Tree amplitudes in gauge theory as scalar MHV diagrams,''
JHEP {\bf 0405}, 070 (2004)
[arXiv:hep-th/0404072];
S.~Gukov, L.~Motl and A.~Neitzke,
``Equivalence of twistor prescriptions for super Yang-Mills,''
arXiv:hep-th/0404085;
J.~B.~Wu and C.~J.~Zhu,
``MHV vertices and scattering amplitudes in gauge theory,''
JHEP {\bf 0407}, 032 (2004)
[arXiv:hep-th/0406085];
I.~Bena, Z.~Bern and D.~A.~Kosower,
``Twistor-space recursive formulation of gauge theory amplitudes,''
arXiv:hep-th/0406133;
J.~B.~Wu and C.~J.~Zhu,
``MHV vertices and fermionic scattering amplitudes in gauge theory with quarks
and gluinos,''
JHEP {\bf 0409}, 063 (2004)
[arXiv:hep-th/0406146];
G.~Georgiou, E.~W.~N.~Glover and V.~V.~Khoze,
``Non-MHV tree amplitudes in gauge theory,''
JHEP {\bf 0407}, 048 (2004)
[arXiv:hep-th/0407027];
X.~Su and J.~B.~Wu,
``Six-quark amplitudes from fermionic MHV vertices,''
arXiv:hep-th/0409228.
}

\lref\structure{
R.~Roiban, M.~Spradlin and A.~Volovich, ``A Googly
Amplitude From The B-Model In Twistor Space,'' JHEP {\bf 0404},
012 (2004) hep-th/0402016; R.~Roiban and A.~Volovich, ``All Googly
Amplitudes From The $B$-Model In Twistor Space,'' hep-th/0402121;
R.~Roiban, M.~Spradlin and A.~Volovich, ``On The Tree-Level
S-Matrix Of Yang-Mills Theory,'' Phys.\ Rev.\ D {\bf 70}, 026009
(2004) hep-th/0403190;
S.~Gukov, L.~Motl and A.~Neitzke,
``Equivalence of twistor prescriptions for super Yang-Mills,''
arXiv:hep-th/0404085;
S.~Giombi, R.~Ricci, D.~Robles-Llana and D.~Trancanelli,
``A note on twistor gravity amplitudes,''
JHEP {\bf 0407}, 059 (2004)
[arXiv:hep-th/0405086];
I.~Bena, Z.~Bern and D.~A.~Kosower,
``Twistor-space recursive formulation of gauge theory amplitudes,''
arXiv:hep-th/0406133.
F.~Cachazo, P.~Svrcek and E.~Witten,
``Twistor space structure of one-loop amplitudes in gauge theory,''
JHEP {\bf 0410}, 074 (2004)
[arXiv:hep-th/0406177].
}

\lref\BenaRY{
I.~Bena, Z.~Bern and D.~A.~Kosower,
``Twistor-space recursive formulation of gauge theory amplitudes,''
arXiv:hep-th/0406133.
}
\lref\looptwistor{
S.~J.~Bidder, N.~E.~J.~Bjerrum-Bohr, L.~J.~Dixon and D.~C.~Dunbar,
``N = 1 supersymmetric one-loop amplitudes and the holomorphic anomaly of
unitarity cuts,''
arXiv:hep-th/0410296;
R.~Britto, F.~Cachazo and B.~Feng,
``Coplanarity in twistor space of N = 4 next-to-MHV one-loop amplitude
coefficients,''
arXiv:hep-th/0411107;
S.~J.~Bidder, N.~E.~J.~Bjerrum-Bohr, D.~C.~Dunbar and W.~B.~Perkins,
``Twistor Space Structure of the Box Coefficients of N=1 One-loop
Amplitudes,''
arXiv:hep-th/0412023.
}

\lref\loopmhv{
A.~Brandhuber, B.~Spence and G.~Travaglini,
``One-loop gauge theory amplitudes in N = 4 super Yang-Mills from MHV
vertices,''
arXiv:hep-th/0407214;
M.~x.~Luo and C.~k.~Wen,
``One-loop maximal helicity violating amplitudes in N = 4 super Yang-Mills
theories,''
JHEP {\bf 0411}, 004 (2004)
[arXiv:hep-th/0410045];
I.~Bena, Z.~Bern, D.~A.~Kosower and R.~Roiban,
``Loops in twistor space,''
arXiv:hep-th/0410054;
M.~x.~Luo and C.~k.~Wen,
``Systematics of one-loop scattering amplitudes in N = 4 super Yang-Mills
theories,''
arXiv:hep-th/0410118;
C.~Quigley and M.~Rozali,
``One-loop MHV amplitudes in supersymmetric gauge theories,''
arXiv:hep-th/0410278;
J.~Bedford, A.~Brandhuber, B.~Spence and G.~Travaglini,
``A twistor approach to one-loop amplitudes in N = 1 supersymmetric Yang-Mills
theory,''
arXiv:hep-th/0410280;
J.~Bedford, A.~Brandhuber, B.~Spence and G.~Travaglini,
``Non-Supersymmetric Loop Amplitudes and MHV Vertices,''
arXiv:hep-th/0412108.
}

\lref\DixonZA{
L.~J.~Dixon, E.~W.~N.~Glover and V.~V.~Khoze,
``MHV rules for Higgs plus multi-gluon amplitudes,''
JHEP {\bf 0412}, 015 (2004)
[arXiv:hep-th/0411092].
}

\lref\BrittoNC{
R.~Britto, F.~Cachazo and B.~Feng,
``Generalized unitarity and one-loop amplitudes in N = 4 super-Yang-Mills,''
arXiv:hep-th/0412103.
}

\lref\BernBT{
Z.~Bern, L.~J.~Dixon and D.~A.~Kosower,
``All Next-to-Maximally-Helicity-Violating One-Loop Gluon Amplitudes in N=4
Super-Yang-Mills Theory,''
arXiv:hep-th/0412210.
}

\lref\RoibanIX{
R.~Roiban, M.~Spradlin and A.~Volovich,
``Dissolving N= 4 loop amplitudes into QCD tree amplitudes,''
arXiv:hep-th/0412265.
}

\lref\ChalmersJB{
G.~Chalmers and W.~Siegel,
``Simplifying algebra in Feynman graphs. II: Spinor helicity from the
spacecone,''
Phys.\ Rev.\ D {\bf 59}, 045013 (1999)
[arXiv:hep-ph/9801220].
}

\lref\KosowerXY{
D.~A.~Kosower,
``Light Cone Recurrence Relations For QCD Amplitudes,''
Nucl.\ Phys.\ B {\bf 335}, 23 (1990).
}

\newsec{Introduction}

Scattering amplitudes of gluons possess a remarkable simplicity
that is not manifest by their computation using Feynman diagrams.

At tree-level the first hints of this hidden simplicity were first
unveiled by the work of Parke and Taylor \parke. They conjectured
a very simple formula for all amplitudes with at most two negative
helicity gluons,\foot{Here we are suppressing a trace factor, a
delta function of momentum conservation and powers of the coupling
constant.}
\eqn\pt{ \eqalign{ & A^{\rm tree}(1^+,2^+,\ldots , n^+) = 0,
\qquad
 A^{\rm tree}(1^-,2^+,\ldots , n^+) = 0, \cr &  A^{\rm tree}(1^-,2^+,\ldots , j^-,\ldots , n^+)
 = {\vev{1~j}^4\over \vev{1~2}\vev{2~3}\ldots \vev{n-1~n}\vev{n~1}}.} }

This formulas were proven by Berends and Giele using their
recursion relations \berandgiele. Many more analytic formulas were obtained
the same way \refs{\mangpxu,\mangparke,\berendsgluon,\KosowerXY}.

Even though these formulas were much simpler than expected, their
form is not as simple as the Parke-Taylor amplitudes.

In a remarkable work \WittenNN, Witten discovered that when
tree-level amplitudes are transformed into twistor space \penrose\
all of them have a simple geometrical description. This led to the
introduction of MHV diagrams (also known as the CSW construction)
in \CachazoKJ, where all tree amplitudes are computed by sewing MHV
amplitudes continued off-shell with Feynman propagators, as well as to a computation using connected instantons that reduces the
problem of finding tree amplitudes to that of solving certain algebraic
equations \RoiSpV. Much
progress has been made in the past year \mhvdiag.

At one-loop, the situation is much more complicated. This is clear
from the fact that the state of the art in QCD is only five-gluon
amplitudes. However, the situation is much better for
supersymmetric amplitudes. Another motivation for studying supersymmetric
amplitudes is that they are useful in the computation of QCD
amplitudes (for a review see \DixonWI). The reason supersymmetric
amplitudes are simpler is because they are four-dimensional cut
constructible \refs{\BernZX,\BernCG}. In particular, $\N=4$ amplitudes have the
simplest structure as they can be written as linear combinations
of scalar box integrals with rational coefficients. Since the
integrals are known explicitly, computing one-loop amplitudes is
reduced to the problem of finding the coefficients.

One-loop $\N=4$ amplitudes are UV finite but IR divergent. The IR
behavior of all one-loop $\N=4$ amplitudes is universal and well understood
\irref. It relates some linear combination of the coefficients to the
tree-level contribution of the amplitude being computed.

These IR equations are usually used as consistency checks of the
coefficients and in many cases as a way of obtaining
hard-to-compute coefficients in terms of other coefficients.
Once the coefficients are computed, they can be used to give new representations of tree-level amplitudes \refs{\BernKY,\BernBT, \RoibanIX}.

However, the situation has changed. In \BrittoNC, we introduced a new
method for computing all coefficients in $\N=4$ one-loop
amplitudes in a simple and systematic manner. Roughly speaking,
every coefficient is given as the product of four tree-level
amplitudes with fewer external legs than the amplitude being
computed.

This leads to a surprising new application of the IR equations.
They now become new recursion relations for tree-level amplitudes!

A particularly simple linear combination of IR equations was found
in \RoibanIX,
\eqn\sim{ A^{\rm tree}_n = {1\over 2}\sum_{i=1}^{n-3}
B_{1,i+1,n-1,n}}
where $B_{abcd}$ denotes the coefficients of a scalar box function
with momenta $K_1 = p_a+p_{a+1}+\ldots + p_{b-1}$, $K_2=p_b+\ldots
+p_{c-1}$, and so on. In \sim\ only two-mass-hard box function
coefficients enter (for $i=1$ and $i=n-3$, the two-mass-hard
becomes a one-mass function).

{}From the result of \BrittoNC, each coefficient $B$ in \sim\ can be 
computed as a sum of products of four tree-level amplitudes of fewer
 external gluons.  The sum runs over all possible particles of the $\N=4$ multiplet and helicity configurations in the loop.

It turns out that one can do better.
In this paper, we propose a new recursion relation for tree-level
amplitudes that involves a sum over terms built
from the product of two tree-level amplitudes times a Feynman
propagator. Schematically, it is given by
\eqn\shcem{ A_n = \sum_{i=1}^{n-3}\sum_{h = \pm}
A^h_{i+2}{1\over P^2_{n,i}}A^{-h}_{n-i} }
where $A_k$ denotes a certain $k$-gluon tree-level amplitude, and
$P_{n,i}$ is the sum of the momenta of gluons $n,1,2,\ldots , i$.
The index $h$ labels the two possible helicity configurations
of the particle being ``exchanged" between two amplitudes.

Note that the form of \shcem\ is quite striking because each term
in the $i$ sum is identical to the factorization limit of $A_n$ in
the $P_{n,i}$ channel. More explicitly, it is well known that the
most leading singular piece of $A_n$ in the kinematical regime
close to $P^2_{n,i}\to 0$ is
\eqn\fact{A_n |_{P^2_{n,i}\to 0} = \sum_{h=\pm}
A^h_{i+2}{1\over P^2_{n,i}}A^{-h}_{n-i}.}
This is known as a multiparticle singularity (for a review see \DixonWI).

Note that in \fact, both tree-amplitudes are on-shell and momentum
conservation is preserved.  This means that each tree amplitude
becomes a physical amplitude.

Very surprisingly, it turns out that the tree amplitudes in
\shcem\ also are on-shell and momentum conservation is preserved.

In order to test our formula \shcem, we recomputed all tree
amplitudes up to seven gluons 
and found complete agreement with the results in the literature. It is
worth mentioning that in \BernKY, formulas for next-to-MHV seven gluon
amplitudes were presented that are simpler than any previously
known form in the literature. Via collinear limits, a very
compact formula for $A(1^-,2^-,3^-,4^+,5^+,6^+)$ was given in \RoibanIX.
Also in \RoibanIX, a very compact formula for the
amplitude $A(1^-,2^-,3^-,4^-,5^+,6^+,7^+,8^+)$ was presented. It
turns out that a straightforward use of our formula \shcem\ gives
rise to the same simple and compact formulas. We also give similar
formulas for all other six-gluon amplitudes.

As a new result we present the eight-gluon amplitude with
alternate helicity configuration, i.e.
$A(1^-,2^+,3^-,4^+,5^-,6^+,7^-,8^+)$.\foot{It would also be possible to derive this amplitude from a suitable limit of the NNMHV amplitudes presented in \DixonZA.} 
We describe how repeated applications of the recursion relations will reduce
any amplitude to a product of three-gluon amplitudes and propagators.  This is very surprising, given that the Yang-Mills Lagrangian has cubic and quartic interactions.
We also discuss an
interesting set of amplitudes that are closed under the recursion
relations and speculate on the possibility of solving for them
explicitly.

This paper is organized as follows: In section 2, we present the
recursion relation in detail.
We illustrate how to use it in practice by giving
 a  detailed calculation of
$A(1^-,2^-,3^-,4^+,5^+,6^+)$. 
In section 3, we present the results
obtained from our recursion relations applied to all amplitudes of up to seven gluons and a particular eight-gluon case. 
In section 4, we present our result for the amplitude $A(1^-,2^+,3^-,4^+,5^-,6^+,7^-,8^+)$.
In
section 5, we discuss some interesting directions for the future.
We give
an outline of a possible proof of our recursion
relations and suggest a way to prove the crucial missing step.
We discuss its possible relation to  MHV diagrams (the CSW construction), and 
point out a 
class of amplitudes closed under the recursion relations.
Finally, in the appendix, we give some details on the calculations
involved in the outline of a possible proof given in section 5.

Throughout the paper, we use the following notation and conventions along with those of \WittenNN.  The external gluon labeled by $i$ carries momentum $p_i$.  
\eqn\ournot{
\eqalign{
 t_i^{[r]} &\equiv (p_i+p_{i+1}+\cdots+p_{i+r-1})^2\ \cr 
\gb{i|\sum_r  p_r |j} &\equiv \sum_r \vev{i~r}[r~j] \cr
\vev{i|(\sum_r p_r )(\sum_s p_s )|j} 
&\equiv \sum_r\sum_s \vev{i~r}[r~s]\vev{s~j} \cr
[i|(\sum_r p_r )(\sum_s p_s )|j] 
&\equiv \sum_r\sum_s [i~r]\vev{r~s}[s~j] \cr
\gb{i|(\sum_r p_r)(\sum_s p_s )(\sum_t p_t )|j} &\equiv 
\sum_r \sum_s \sum_t \vev{i~r}[r~s]\vev{s~t}[t~j] 
}}

\newsec{Recursion Relations}

Consider any $n$-gluon tree-level amplitude with any helicity
configuration. Without loss of generality let us take the labels
of the gluons such that the $(n-1)$-th gluon has negative helicity
and the $n$-th gluon has positive helicity.\foot{Recall that
amplitudes with all positive or all negative helicity gluons vanish
\refs{\parke,\berandgiele}.} Then we claim that the following recursion relation for
tree amplitudes is valid:
\eqn\retrue{\eqalign{ & A_n(1,2,\ldots , (n-1)^-,n^+) = \cr &
\sum_{i=1}^{n-3}\sum_{h=+,-} \left ( A_{i+2}({\hat n},1,2,\ldots
i,-{\hat P}^h_{n,i} ) {1\over P^2_{n,i}} A_{n-i}(+{\hat
P}^{-h}_{n,i}, i+1,\ldots , n-2, {\hat{n-1} } ) \right), } }
where
\eqn\deff{ \eqalign{ P_{n,i} & = p_n+p_1+\ldots + p_i, \cr \hat
P_{n,i} & = P_{n,i} +{P_{n,i}^2\over \gb{n-1|P_{n,i}|n}}
\lambda_{n-1} \tilde\lambda_{n}, \cr \hat p_{n-1} & =  p_{n-1}
-{P_{n,i}^2\over \gb{n-1|P_{n,i}|n}} \lambda_{n-1}
\tilde\lambda_{n} , \cr \hat p_{n} & =  p_{n} +{P_{n,i}^2\over
\gb{n-1|P_{n,i}|n}} \lambda_{n-1} \tilde\lambda_{n}. } }

At this point we should make some comments about this formula.
First, note that the momenta $P_{n,i}$, $p_{n-1}$ and $p_{n}$ are all shifted in
the same way. The term we add,
\eqn\add{{P_{n,i}^2\over \gb{n-1|P_{n,i}|n}} \lambda_{n-1}
\tilde\lambda_{n}, }
seems peculiar at this point, but it arises naturally from the
discussion in section 5. The shift \add\ is not parity invariant. Moreover,
it cannot be interpreted as a vector in Minkowski space, since $\lt$ and $\lambda$ are independent,
but it has
a natural meaning in $(--++)$ signature, as we discuss in section
5.

Note that each tree-level amplitude in \retrue\ has all external
gluons on-shell. Indeed, it is easy to see that ${\hat P}_{n,i}^2
= {\hat p}^2_n = {\hat p}^2_{n-1} = 0$. It is interesting to note
that ${\hat P}_{n,i}$ is a generalization of the formula used in
\refs{\KosowerYZ,\BenaRY} to define non-MHV amplitudes off-shell. In contrast, here we
used it in order to keep the amplitudes on-shell while momentum
conservation is preserved.

\ifig\kkmbp{Pictorial representation of the recursion relation
\retrue. Note that the difference between the terms in the two
sums is just the helicity assignment of the internal line.}
{\epsfxsize=0.85\hsize\epsfbox{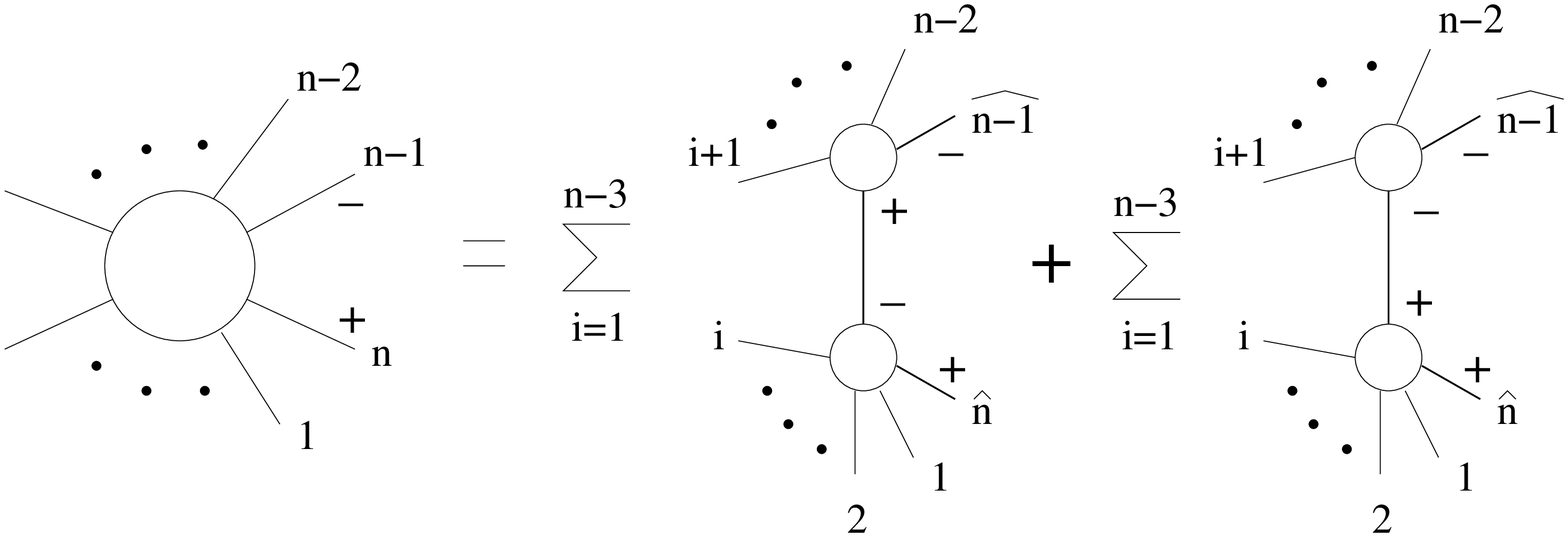}}

The fact that momentum conservation is preserved in each of the
tree-level amplitudes, as anticipated in the introduction, might
be a little puzzling at first. Consider in particular the
limiting cases in the sum \retrue, i.e, $i=1$ and $i=n-3$. For these
two values of $i$, one tree amplitude is a three-gluon amplitude
on-shell, and it is well known that this vanishes in Minkowski
signature $(-+++)$. Here, on the other hand, the momenta are
taken in $(--++)$ signature, and three-gluon amplitudes on-shell
are non-trivial.

\subsec{Tips for Using the Recursion Relation}

Let us explain the way we use \retrue\ and \deff\ in practice.
First note that from the definition of $\hat p_{n-1}$ and $\hat
p_{n}$ one can read off their spinor components very easily.
Recall that $p_{n-1} =\lambda_{n-1}\tilde\lambda_{n-1}$ and
$p_n=\lambda_{n}\tilde\lambda_{n}$, therefore
\eqn\reado{ \eqalign{ \lambda_{\hat{n-1}} & = \lambda_{n-1}, \cr
\tilde\lambda_{\hat{n-1}} & =  \tilde\lambda_{n-1}
-{P_{n,i}^2\over \gb{n-1|P_{n,i}|n}} \tilde\lambda_{n}, \cr
\lambda_{\hat{n}} & = \lambda_{n} +{P_{n,i}^2\over
\gb{n-1|P_{n,i}|n}} \lambda_{n-1}, \cr \tilde\lambda_{\hat{n}} & =
\tilde\lambda_{n}.   } }
Finally, we use the following identities to compute any spinor
product involving ${\hat P}_{n,i}$.
\eqn\jull{ \eqalign{  \vev{\bullet~ {\hat P}_{n,i} } & =
-\gb{\bullet | P_{n,i} | n }\times {1\over \omega} \cr  [{\hat
P}_{n,i}~\bullet ] & = -\gb{ n-1| P_{n,i}| \bullet }\times {1\over
\bar\omega} } }
Here $\omega = [ {\hat P}_{n,i}~~ n]$ and $\bar\omega = \vev{n-1~~
{\hat P}_{n,i}}$. Since the $\hat P_{n,i}$ have opposite
helicities on both amplitudes, the product $A_{i+2}A_{n-i}$
must have degree zero under the rescaling of $\lambda_P\to
t\lambda_P$ and $\tilde\lambda_P \to t^{-1}\tilde\lambda_P$.
Therefore, the factors $\omega$ and $\bar\omega$ can only show up
in the final answer in the invariant combination
$\omega\bar\omega$. This combination is easy to compute and it is
given by $\gb{ n-1| P_{n,i} | n} $.

In practice, it is very useful to note the following.
Due to our choice of reference spinors in
\add, one can easily show that diagrams with an upper vertex of
the form $(++-)$ or with a lower vertex of the form $(--+)$
vanish. For example, assume that the $(n-2)$-th gluon has positive
helicity.  Then the first term in figure 1 vanishes for $i=n-3$, 
i.e., the term
with upper vertex $((n-2)^+, \hat{n-1}^-, {\hat P}^+_{n,n-3})$.

We illustrate this procedure in detail with calculation of a
six-gluon amplitude below; the same procedure was used 
 in the computation of all the results in sections 3 and 4.

\subsec{An Explicit Example}

As a first application of our formula, we compute the next-to-MHV
six-gluon amplitude $A(1^-,2^-,3^-,4^+,5^+,6^+)$. Note that we
have shifted the labels with respect to the conventions in the
previous section. This is done in order to compare more easily to the
result to the known formula in the literature \refs{\mangpxu, \mangparke}.

Here we choose the reference gluons to be $\hat 3$ and $\hat 4$.
There are three possible configurations of external gluons. Only
one helicity configuration for the internal gluon gives a nonzero
answer. 
\ifig\kkmbp{Configurations contributing to the six-gluon amplitude
$A_6(1^-,2^-,3^-,4^+,5^+,6^+)$. Note that $(a)$ and $(c)$ are
related by a flip and a conjugation. $(b)$ vanishes for either helicity configuration of the internal line.}
{\epsfxsize=0.70\hsize\epsfbox{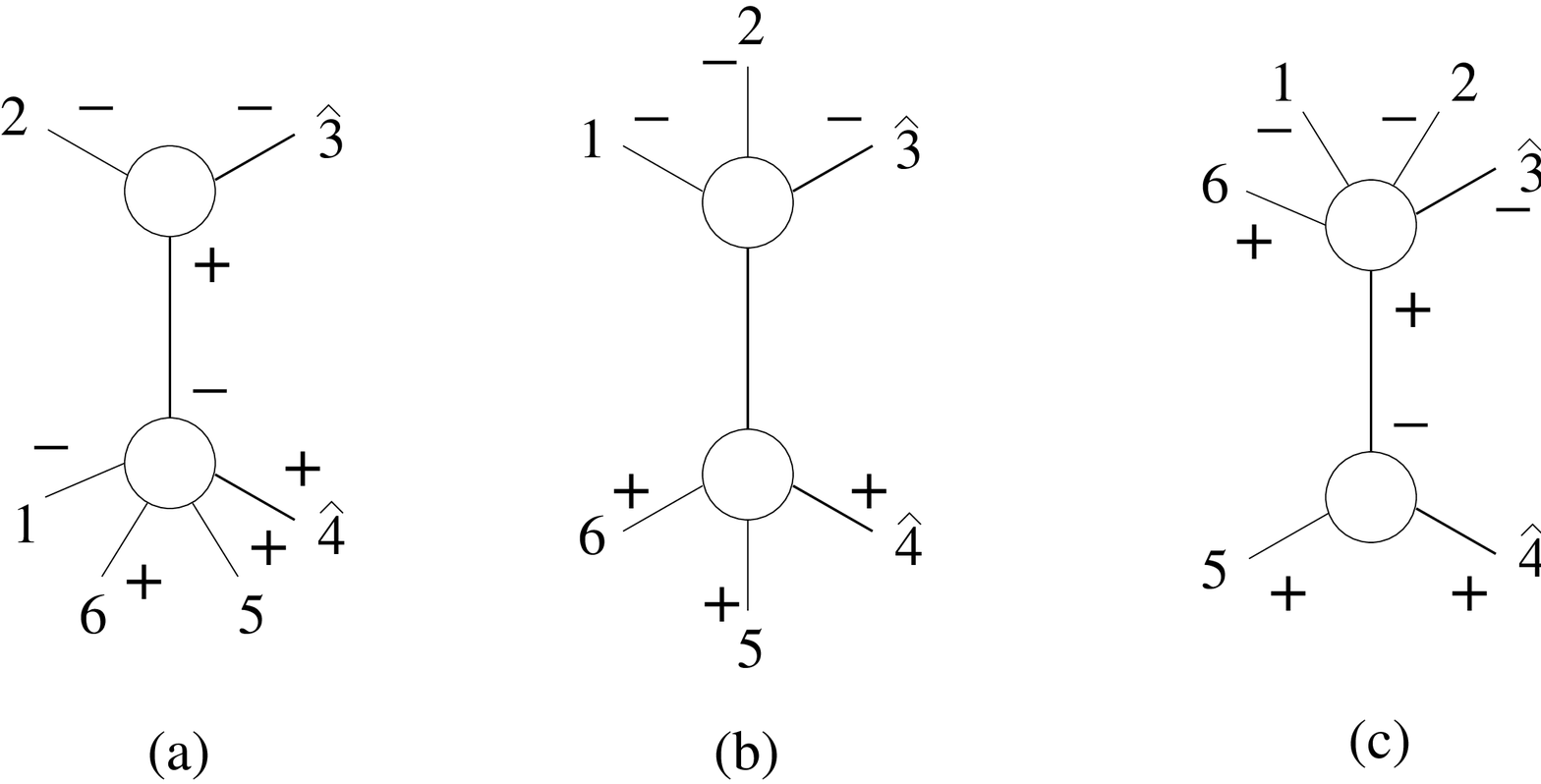}}
This is shown in \kkmbp. Note that for this helicity
configuration, the middle graph vanishes. Therefore, we are left
with only two graphs to evaluate. Moreover, the two graphs are
related by a flip of indices composed with a conjugation.
Therefore, only one computation is needed.

Let us compute in detail the contribution coming from the first
graph shown in \kkmbp(a). The contribution of this term is given
by the product of two MHV amplitudes times a propagator,
\eqn\conja{ \left( \vev{2~\hat 3}^3\over \vev{\hat 3 ~\hat
P}\vev{\hat P ~ 2} \right) {1\over t_2^{[2]}}\left( {\vev{1~\hat
P}^3 \over \vev{\hat P ~ \hat 4}\vev{\hat 4~ 5}\vev{5~6}\vev{6~1}
} \right). }
This formula can be simplified by noting that
\eqn\holu{ \eqalign{  \lambda_{\hat 3} & = \lambda_3, \cr
\lambda_{\hat 4} & = \lambda_4 - {t_2^{[2]}\over
\vev{3~2}[2~4]}\lambda_3, \cr \vev{\bullet ~ \hat P} & = -{ \gb{
\bullet | 2+3 | 4} \over [\hat P ~ 4]}. }}
Using \holu\ it is straightforward to find \conja\
\eqn\lala{ {\gb{ 1 | 2+3 | 4}^3 \over [2~3][3~4]\vev{5~6}\vev{6~1}
t_2^{[3]}\gb{ 5 | 3+4 | 2} } .}

Finally, applying the flip that takes $i\to i+3$ and a
conjugation, i.e., $\vev{~~} \leftrightarrow [~~]$,
to \lala\ we find the contribution from configuration $(c)$.
Adding both contributions and factoring out a common term, we get
\eqn\allsix{ A(1^-,2^-,3^-,4^+,5^+,6^+) = {1\over \gb{ 5 | 3+4 |
2}}\left( {\gb{ 1 | 2+3 | 4}^3 \over [2~3][3~4]\vev{5~6}\vev{6~1}
t_2^{[3]}} + {\gb{ 3 | 4+5 | 6}^3 \over
[6~1][1~2]\vev{3~4}\vev{4~5} t_3^{[3]} }\right).}

Quite surprisingly, this is the formula found in \RoibanIX\ by taking a
collinear limit of a seven-gluon amplitude representation given in
\BernKY.

\newsec{Previously Known Amplitudes}

In this section we recompute all known tree-level amplitudes of
gluons for $n\leq 7$ using our recursion relations. It turns out
that all the formulas we get  come out naturally in a very compact
form. We start with the MHV amplitudes and show that they satisfy
the recursion relation. For next-to-MHV amplitudes, we compute all
six- \refs{\mangpxu,\mangparke} and seven-gluon amplitudes \berendsgluon. Finally, we compute the
next-to-next-to-MHV eight gluon amplitude with four adjacent
minuses \RoibanIX.

All the results presented in this section and the next  were
computed using exactly the same technique as in the example of
section 2.2. Here we will not repeat the details but we will
indicate explicitly the contribution coming from each term in
\retrue. It turns out that in all the cases considered here only
one helicity configuration of the internal propagator gives a
non-zero contribution. Therefore, in order to specify a given term
in \retrue, it is enough to give the gluons in each tree amplitude
and the reference gluons denoted by a hat. In the example of section
2.2, we would refer to the contribution from \kkmbp(a) as
$(2,\hat 3|\hat 4,5,6,1)$ and from \kkmbp(c) as $(6,1,2,\hat
3|\hat 4, 5)$.

\subsec{MHV Amplitudes}

We  show here that the Parke-Taylor formula for MHV amplitudes
\pt\ satisfies the recursion relations.

Consider the amplitude $A(1^-,2^+,\ldots ,(j-1)^+,
j^-,(j+1)^+,\ldots ,(n-1)^+, n^+ )$ and assume that \pt\ is valid
for all MHV amplitudes with fewer than $n$ gluons.\foot{Of course,
we also assume that the conjugate amplitudes for fewer than $n$
gluons are also valid.} Using the recursion relation \retrue, with 1 and 2 chosen as the reference gluons, we
find that only one term is non-zero. It is the term given by
$(4,5,\ldots,n,\hat 1~|~\hat 2,3)$. Its contribution is given
by
\eqn\mhvre{ {\vev{j~\hat 1}^4\over \vev{4~5}\ldots
\vev{n-1~n}\vev{n~\hat 1}\vev{\hat 1~\hat P}\vev{\hat P~ 4}}\times
{1\over \vev{2~3}[2~3]}\times {[\hat 2~ 3]^3\over [3~\hat P][\hat
P~ \hat 2]}, }
where $P = p_2+p_3$.

After using \reado\ and \jull\ to remove the hats, we find
\eqn\jipp{\eqalign{ & A(1^-,2^+,\ldots ,(j-1)^+,
j^-,(j+1)^+,\ldots ,(n-1)^+, n^+ ) = \cr & {\vev{1~j}^4\over
\vev{4~5}\vev{5~6}\ldots\vev{n-1~n}\vev{n~1}\vev{1~2}\vev{2~3}\vev{3~4}}.}}

\bigskip

\subsec{Six-Gluon Amplitudes}

We now compute all next-to-MHV six-gluon amplitudes.

The case with three adjacent minus helicities was presented in
detail in the previous section, and the answer appears in \allsix.

The next configuration gives a three-term expression.  These three
terms are the contributions from $(2,\hat{3}|\hat{4},5,6,1)$, $(1,2,\hat{3}|\hat{4},5,6)$, and $(6,1,2,\hat{3}|\hat{4},5)$ respectively.
\eqn\SixMone{\eqalign{ A(1^+,2^+,3^-,4^+,5^-,6^-) & =
{[2~4]^4\vev{5~6}^3\over [2~3][3~4] \vev{6~1} t_{2}^{[3]}
\gb{1|2+3|4} \gb{5|3+4|2}}\cr & + { \gb{3|1+2|4}^4\over
\vev{1~2}\vev{2~3}[4~5][5~6]  t_{1}^{[3]} \gb{1|2+3|4}
\gb{3|1+2|6}} \cr & + { [1~2]^3 \vev{3~5}^4\over
[6~1]\vev{3~4}\vev{4~5}  t_{3}^{[3]} \gb{5|3+4|2} \gb{3|4+5|6}}.
}}
Similarly, for the final configuration of a next-to-MHV six-gluon
amplitude, there are three terms that are the contributions from
$(1,\hat{2}|\hat{3},4,5,6)$, $(6,1,\hat{2}|\hat{3},4,5)$, and  $(5,6,1,\hat{2}|\hat{3},4)$ respectively.
\eqn\SixMtwo{\eqalign{ A(1^+,2^-,3^+,4^-,5^+,6^-) & = { [1~3]^4
\vev{4~6}^4 \over [1~2] [2~3] \vev{4~5} \vev{5~6} t_{1}^{[3]}
\gb{6|1+2|3} \gb{4|2+3|1}} \cr & + { \vev{2~6}^4 [3~5]^4\over
\vev{6~1}\vev{1~2} [3~4] [4~5] t_{3}^{[3]} \gb{6|4+5|3}
\gb{2|3+4|5}} \cr & + { [1~5]^4 \vev{2~4}^4 \over
\vev{2~3}\vev{3~4} [5~6] [6~1] t_{2}^{[3]} \gb{4|2+3|1}
\gb{2|3+4|5}}. }}
These two expressions have been checked against the known results \mangpxu. It is amazing to notice that for
$A(1^+,2^-,3^+,4^-,5^+,6^-)$, the second two terms can be obtained from
the first
 by  shifting all indices by $i\to i+2$ and $i\to i+4$.

\subsec{Seven-Gluon Amplitudes}

Now we use our recursion relation to calculate the tree level
next-to-MHV amplitude of seven gluons and compare with results
given in \BernKY.  We follow the conventions of that paper to write the four independent helicity configurations.

For configuration A:$(1^-,2^-,3^-,4^+,5^+,6^+,7^+)$, there are only two nonzero contributions, namely from $(2,\hat{3}|\hat{4},5,6,7,1)$ and $(6,7,1,2,\hat{3}|\hat{4},5)$.  The first involves only MHV amplitudes, so it is just one term.
The second involves the next-to-MHV six-gluon amplitude with two terms.  We write these three terms in order here:
\eqn\SevenA{\eqalign{& A(1^-,2^-,3^-,4^+,5^+,6^+,7^+) = \cr 
& {\gb{1|2+3|4}^3 \over t_2^{[3]}\vev{5~6}\vev{6~7}\vev{7~1}[2~3][3~4]
\gb{5|4+3|2}} \cr &
-{1 \over \vev{3~4}\vev{4~5}\gb{6|7+1|2}}
\left( {\vev{3|(4+5)(6+7)|1}^3 \over t_3^{[3]}t_6^{[3]}\vev{6~7}\vev{7~1}\gb{5|4+3|2}} + {\gb{3|2+1|7}^3 \over t_7^{[3]} \vev{6~5} [7~1][1~2]} \right). \cr
}}
Term by term, this expression is equal to 
$c_B +c_{347}|_{\rm flip}+c_{347}$ from \BernKY, which is exactly the compact formula given there.
For configuration B:$(1^-,2^-,3^+,4^-,5^+,6^+,7^+)$, there are
three nonzero contributions. We write the formula in the following
order:  the single term from $(3,\hat{4}|\hat{5},6,7,1,2)$, the single
term from $(2,3,\hat{4}|\hat{5},6,7,1)$, and the three terms from
$(7,1,2,3,\hat{4}|\hat{5},6)$.
\eqn\SevenB{\eqalign{ & A(1^-,2^-,3^+,4^-,5^+,6^+,7^+) = \cr &
{\vev{1~2}^3 [3~5]^4\over t_{3}^{[3]} [3~4][4~5]
\vev{6~7}\vev{7~1}\gb{2|3+4|5}\gb{6|4+5|3}}\cr &+ {\vev{2~4}^4
\gb{1|7+6|5}^3 \over  t_{2}^{[3]}t_{6}^{[3]}
\vev{2~3}\vev{3~4}\vev{6~7}\vev{7~1}\gb{2|3+4|5}
\vev{6|(7+1)(2+3)|4}} \cr
 &+ {\vev{1~2}^3 \gb{4|5+6|3}^4 \over
 t_{4}^{[3]} t_{7}^{[3]} \vev{4~5}\vev{5~6}\vev{7~1}
\gb{6|4+5|3} \gb{7|1+2|3} \vev{4|(5+6)(7+1)|2}} \cr & + {
\gb{4|1+2|3}^4\over  t_{1}^{[3]} [1~2][2~3]\vev{4~5}\vev{5~6}
\vev{6~7}\gb{4|3+2|1}\gb{7|1+2|3}} \cr & + {\vev{2~4}^4
\gb{4|5+6|7}^3 \over \vev{2~3}\vev{3~4}\vev{4~5}
\vev{5~6}[7~1]\gb{4|3+2|1} \vev{4|(5+6)(7+1)|2}
\vev{6|(7+1)(2+3)|4}}. }}
Term by term, this expression is equal to
$c_{145}+c_A+c_E+c_{236}+c_{136}$ from \BernKY.  
This is not the exact compact formula given in that paper, but
it is possible to
derive from the relations given there that this is the correct
tree amplitude.
It would be interesting to check if a different choice of reference gluons reproduces the compact form of \BernKY, term by term.

For configuration C:$(1^-,2^-,3^+,4^+,5^-,6^+,7^+)$, there are
four nonzero contributions. We write the formula in the following
order:  the single term from $(1,\hat{2}|\hat{3},4,5,6,7)$, the single
term from $(7,1,\hat{2}|\hat{3},4,5,6)$, the single term from
$(6,7,1,\hat{2}|\hat{3},4,5)$, and the three terms from
$(5,6,7,1,\hat{2}|\hat{3},4)$. \eqn\SevenC{\eqalign{ &
A(1^-,2^-,3^+,4^+,5^-,6^+,7^+) = \cr & {\gb{5|1+2|3}^4\over
t_{1}^{[3]} [1~2][2~3]\vev{4~5} \vev{5~6}\vev{6~7}\gb{7|1+2|3}
\gb{4|3+2|1}} \cr &+ { \vev{1~2}^3 \gb{5|4+6|3}^4 \over
t_{4}^{[3]}
 t_{7}^{[3]} \vev{4~5}\vev{5~6}\vev{7~1}\gb{6|5+4|3}
\gb{7|1+2|3} \vev{4|(5+6)(7+1)|2}} \cr &+ { \vev{1~2}^3 [3~4]^3
\over  t_{3}^{[3]} [4~5]\vev{6~7}\vev{7~1}\gb{2|3+4|5}
\gb{6|4+5|3}} \cr
  & +{ \vev{1~2}^3\gb{2|3+4|6}^4 \over \vev{7~1}
\vev{2~3}\vev{3~4}[5~6] \gb{2|1+7|6} \gb{2|3+4|5}
\vev{2|(3+4)(5+6)|7} \vev{4|(5+6)(7+1)|2}} \cr & 
+ {\vev{2|(3+4)(7+6)|5}^4 \over
t_{2}^{[3]}t_{5}^{[3]}\vev{2~3}\vev{3~4}
\vev{5~6}\vev{6~7}\gb{5|6+7|1} \gb{4|2+3|1} \vev{2|(3+4)(5+6)|7}}
\cr & + { \vev{2~5}^4 [6~7]^3\over
t_{6}^{[3]}\vev{2~3}\vev{3~4}\vev{4~5} [7~1]\gb{5|6+7|1}
\gb{2|7+1|6}}. }} Term by term, this expression is equal to
$c_{236}+c_A|_{\rm flip}+c_{367}|_{\rm flip}+c_{357}|_{\rm
flip}+c_{C}|_{\rm flip}+c_{147}$ from \BernKY.  It is possible to
derive from the formulas in that paper that this is the correct
tree amplitude.

For configuration D:$(1^-,2^+,3^-,4^+,5^-,6^+,7^+)$, there are
again four nonzero contributions. We write the formula in the
following order:  the single term from $(2,\hat{3}|\hat{4},5,6,7,1)$,
the single term from $(1,2,\hat{3}|\hat{4},5,6,7)$, the single term
from$(7,1,2,\hat{3}|\hat{4},5,6)$,  and the three terms from
$(6,7,1,2,\hat{3}|\hat{4},5)$.
\eqn\SevenD{\eqalign{ & A(1^-,2^+,3^-,4^+,5^-,6^+,7^+) = \cr &
{\vev{1~5}^4 [2~4]^4 \over t_{2}^{[3]}[2~3][3~4]
\vev{5~6}\vev{6~7}\vev{7~1}\gb{1|2+3|4}\gb{5|3+4|2} }\cr &
-{\vev{1~3}^4 \gb{5|6+7|4}^4 \over t_{1}^{[3]}t_{5}^{[3]}
\vev{1~2}\vev{2~3}\vev{5~6}\vev{6~7}\gb{1|2+3|4} \gb{7|5+6|4}
\vev{5|(6+7)(1+2)|3}}\cr &+ {\vev{1~3}^4 [4~6]^4 \over
t_{4}^{[3]} \vev{7~1}\vev{1~2}\vev{2~3}[4~5][5~6]\gb{3|4+5|6}
\gb{7|5+6|4}}\cr
 & -{ \vev{3~5}^4 [2~7]^4 \over
t_{7}^{[3]} \vev{3~4}\vev{4~5}\vev{5~6}[7~1][1~2] \gb{6|7+1|2}
\gb{3|2+1|7}} \cr & +{\vev{3~5}^4 \gb{1|6+7|2}^4\over
t_{3}^{[3]}t_{6}^{[3]} \vev{6~7}\vev{7~1}\vev{3~4}\vev{4~5}
\gb{6|7+1|2}\gb{5|3+4|2} \vev{3|(4+5)(6+7)|1}}\cr & -{ \vev{1~3}^4
\vev{3~5}^4 [6~7]^3 \over \vev{1~2}\vev{2~3}
\vev{3~4}\vev{4~5}\gb{3|4+5|6} \gb{3|2+1|7}
\vev{3|(4+5)(6+7)|1}\vev{5|(6+7)(1+2)|3}}. } }
Term by term, this expression is equal to
$c_{347}+c_B+c_{256}+c_{256}|_{\rm flip}+c_{B}|_{\rm
flip}+c_{257}$ from \BernKY. This is exactly the same compact
formula for the tree amplitude given in that paper.

\subsec{Eight-Gluon Amplitude}

A very compact formula for the next-to-next-to-MHV amplitude
$A(1^-,2^-,3^-,4^-,5^+,6^+,7^+,8^+)$   was computed very
recently in \RoibanIX.

Using our formula \retrue\ it is easy to see that there are only
two terms contributing to the amplitude. They are
$(3,\widehat{4}|\widehat{5},6,7,8,1,2)$ and
$(7,8,1,2,3,\widehat{4}|\widehat{5},6)$. Let us first consider the
contribution $(3,\widehat{4}|\widehat{5},6,7,8,1,2)$. Using the
seven-gluon amplitude \SevenA,
we find immediately that
\eqn\eighta{\eqalign{  I_a & = { \gb{1|K_{2}^{[3]}| 5}^3 \over
t_{2}^{[4]} [2~3][3~4][4~5] \vev{6~7} \vev{7~8}\vev{8~1}
\gb{6|K_{3}^{[3]}|2} } \cr & -{ \gb{ 1|K_{7}^{[2]}
K_{3}^{[4]}K_{3}^{[2]}|5}^3 \over t_{7}^{[3]} t_{3}^{[4]}
t_{3}^{[3]} [3~4][4~5]\vev{7~8}\vev{8~1} \gb{6|K_{4}^{[2]}|3}
\gb{7|K_{8}^{[2]}|2} \gb{6|K_{3}^{[3]}|2}}\cr & - {
[8|K_{1}^{[2]}K_{3}^{[2]}|5]^3 \over \vev{7~6}[8~1][1~2][3~4][4~5]
t_{3}^{[3]}t_{8}^{[3]} \gb{7|K_{8}^{[2]}|2} \gb{6|K_{4}^{[2]}|3}},
}}
where the order of these three terms is the same as in \SevenA.
Our convention here is that $K_i^{[r]} \equiv (p_i+p_{i+1}+\ldots+p_{i+r-1})$.
The contribution from $(7,8,1,2,3,\widehat{4}|\widehat{5},6)$ is just
the flip of $I_a$ by relabeling $i\rightarrow (9-i)$ and
exchanging $\vev{~}$ and $[~]$.

Let us compare our result with the one given in equation (1) of \RoibanIX. 
It is easy to see that the first two
terms match,
 while the last term in \eighta\ is related by the above flip to the term 
written explicitly in \RoibanIX.


\newsec{Result for $A(1^+,2^-,3^+,4^-,5^+,6^-,7^+,8^-)$  }

In this section we present the NNMHV eight-gluon amplitude with
alternating helicities. There are five different configurations of
gluons we have to consider.  Here again, to save space, we use the notation
$K_i^{[r]} \equiv (p_i+p_{i+1}+\ldots+p_{i+r-1})$.

First we find the contribution of $(1,\hat{2}|\hat{3},4,5,6,7,8)$.  It is
\eqn\Eighttwosix{\eqalign{ I_1 &= {[1~3]^4\vev{4~6}^4\vev{6~8}^4
 \over [1~2][2~3]\vev{4~5}\vev{5~6}\vev{6~7}\vev{7~8}
\gb{6|K_{7}^{[2]}|1} \gb{6|K_{4}^{[2]}|3}
\vev{6|K_{7}^{[2]}K_{1}^{[3]}|4} \vev{8|K_{1}^{[3]}K_{4}^{[2]}|6}}
\cr & 
+{ [1~3]^4 [5~7]^4 \vev{4~8}^4 \over [1~2][2~3]
[5~6][6~7] t_{1}^{[3]} t_{5}^{[3]}\gb{4|K_{2}^{[2]}|1}\gb{4|K_{5}^{[2]}|7} 
\gb{8|K_{1}^{[2]}|3}\gb{8|K_{6}^{[2]}|5}}\cr & 
- { [1~3]^4
\vev{4~6}^4 \gb{8|K_{1}^{[3]}|7}^4 \over [1~2][2~3]\vev{4~5}\vev{5~6}
t_{1}^{[3]}t_{4}^{[3]}t_{8}^{[4]}
\gb{4|K_{5}^{[2]}|7}\gb{8|K_{1}^{[2]}|3} [1|K_{2}^{[2]}
K_{4}^{[3]}|7] \vev{8|K_{1}^{[3]}K_{4}^{[2]}|6} } \cr & 
+ {
[1~3]^4 [7~1]^4 \vev{4~6}^4\over [1~2][2~3]\vev{4~5}\vev{5~6}[7~8][8~1]\gb{4|K_{2}^{[2]}|1}\gb{6|K_{7}^{[2]}|1} [1|K_{2}^{[2]}
K_{4}^{[3]}|7] [3|K_{4}^{[3]}K_{7}^{[2]}|1]
 }\cr & 
-{ [1~3]^4 \vev{6~8}^4
\gb{4|K_{1}^{[3]}|5}^4 \over [1~2][2~3]\vev{6~7}\vev{7~8} t_{1}^{[3]} t_{6}^{[3]}t_{5}^{[4]}
\gb{4|K_{2}^{[2]}|1}\gb{8|K_{6}^{[2]}|5}
 [5|K_{6}^{[3]}K_{1}^{[2]}|3]\vev{6|K_{7}^{[2]}K_{1}^{[3]}|4}}
\cr & 
+ { [1~3]^4 [3~5]^4\vev{6~8}^4
 \over [1~2][2~3][3~4][4~5]\vev{6~7}\vev{7~8}\gb{6|K_{4}^{[2]}|3} \gb{8|K_{1}^{[2]}|3} [3|K_{4}^{[2]}K_{6}^{[3]}|1]
 [5|K_{6}^{[3]}K_{1}^{[2]}|3]
 }. }}
{}From this one we can get the contribution of
$(5,6,7,8,1,\hat{2}|\hat{3},4)$ by 
a conjugate flip operation that
exchanges indices
$2\leftrightarrow 3$, $1\leftrightarrow 4$,  $8\leftrightarrow 5$,
 $7\leftrightarrow 6$ and $\vev{~~}\leftrightarrow [~~]$.

Second, we get that  the  contribution of $(8,1,\hat{2}|\hat{3},4,5,6,7)$
is
\eqn\Eightthreefive{\eqalign{ I_2 &= -{\vev{8~2}^4  \over
\vev{8~1} \vev{1~2} \gb{8| K_{8}^{[3]} |3} t_{8}^{[3]}} \left(
{[3~5]^4 \gb{6| K_{8}^{[3]}|3}^4 \over [3~4][4~5] \vev{6~7} \gb{7|
K_{8}^{[3]}|3} \gb{6|K_{4}^{[2]}|3}
\gb{2|K_{8}^{[3]}K_{6}^{[2]}K_{4}^{[2]}|3}
[5|K_{6}^{[2]}K_{8}^{[3]}|3]}\right. \cr & + { [5~7]^4
\gb{4|K_{8}^{[3]}|3}^4 \over [5~6][6~7]  t_{8}^{[4]}
 t_{5}^{[3]} \vev{4|K_{5}^{[3]}K_{8}^{[3]}|2}
[5|K_{5}^{[3]}K_{8}^{[3]}|3] \gb{4|K_{5}^{[3]}|7}} \cr & \left. -{
[3~7]^4\vev{4~6}^4\over \vev{4~5} \vev{5~6}
 t_{4}^{[3]} \gb{2|K_{8}^{[3]}|7} \gb{6|K_{4}^{[3]}|3}
\gb{4|K_{4}^{[3]}|7}}\right). }}
{}From this expression, by the same conjugate flip, we can get the
contribution from $(6,7,8,1,\hat{2}|\hat{3},4,5)$.

Finally, from   $(7,8,1,\hat{2}|\hat{3},4,5,6)$ we get
\eqn\Eightfourfour{\eqalign{ I_3 & = -{\vev{8~2}^4 \over
\vev{7~8}\vev{8~1}\vev{1~2} \gb{7|K_{7}^{[4]}|3}}
 {1\over t_{7}^{[4]} }
{ [3~5]^4\over [3~4][4~5][5~6] \gb{2|K_{3}^{[4]}|6}} \cr & - {
[7~1]^4 \gb {2|K_{7}^{[4]}|3}^2 \over [7~8][8~1] t_{7}^{[3]}
[3|K_{3}^{[4]}K_{7}^{[2]}|1] \gb{2|K_{7}^{[4]}|7}}
 {1\over t_{7}^{[4]} }
{ \vev{4~6}^4 \gb{2|K_{3}^{[4]}|3}^2 \over
\vev{4~5}\vev{5~6}t_{4}^{[3]} \vev{4|K_{5}^{[2]}K_{7}^{[4]}|2}
\gb{6|K_{3}^{[4]}|3}}. }}
Notice that there are two terms corresponding to two different
helicity assignments on the propagator.

The final result is
\eqn\Eightfinal{ A(1^+,2^-,3^+,4^-,5^+,6^-,7^+,8^-) =[I_1+I_1^{\rm
conj. flip}]+[I_2+I_2^{\rm conj. flip}]+ I_3. }
%
\subsec{Symmetric Form}

The amplitude $A(1^+,2^-,3^+,4^-,5^+,6^-,7^+,8^-)$ has a high degree of symmetry.  Our recursion procedure breaks almost all the symmetry by choosing two reference gluons.  Here we show that from the formula above, we can in fact deduce a fully symmetric expression.

  The full symmetry group is the dihedral group of order 16, with the following two generators acting on the gluon indices:
\eqn\dihedral{ g:~i \rightarrow i+1~{\rm with~conjugation}, \qquad r:~i \rightarrow -i \quad ({\rm mod}~8). }
These satisfy the relations $g^8=1, r^2=1$ and $rgrg=1$.  Many of the terms of  \Eighttwosix, \Eightthreefive, and \Eightfourfour\ are related by these symmetries.  Thus it is possible to represent those equations, term by term, as follows:
\eqn\familterm{\eqalign{
I_1 & = T + U + V + g^3 T + g^5 V + g^5 T \cr
I_2 & = W + g^4 V + g^3 U \cr
I_3 &=  X + g^3 V. 
}}

The five independent terms we have defined have the following symmetry:
\eqn\papaya{
T = g^4 r T, \quad U = g^4 U = r U, \quad V = g^7 r V, \quad X = g^5 r X.
}

The flip symmetry used in \Eightfinal\ is $g^5r$ in this notation.  
After performing this action on $I_1$ and $I_2$ to get the amplitude \Eightfinal and using the relations in \papaya, we can reorder the terms to get the expression
\eqn\marching{\eqalign{
 A(1^+,2^-,& 3^+,4^-,5^+,6^-,7^+,8^-) =
 (T + g T + g^3 T + g^4 T + g^5 T +g^6 T ) \cr
& + (U+ gU+ g^2 U + g^3 U) 
 + (V+ gV  + g^2 V + g^3 V  + g^4 V + g^5 V + g^6 V) \cr
& +(W + g^5 r W) + X.
 }}
We have checked that the last three terms, $(W + g^5 r W) + X$, are equal to 
the sum $g^2 T + g^7 T + g^7 V$.  Making this substitution in \marching\ gives an expression with all the required symmetry manifest. The amplitude is the sum of the orbits of the terms $T, U$ and $V$.


\newsec{Future Directions}

In this section we present some of the future directions that are
natural to explore given the success of the recursion relation
\retrue.

First we give an outline of a possible proof of the recursion relation
\retrue. Then we show how one can use the recursion relation
several times to write any amplitude as the sum of terms computed
from only trivalent vertices with helicities $(++-)$ and $(--+)$.
These new diagrams hint at a connection to a string theory whose target space is the Quadric.
We also comment on a possible connection to MHV diagrams. We give
a set of amplitudes that are closed under the recursion relations
and suggest that one can hope to solve it explicitly. Finally, we
comment on a possible extension of the recursion relations that
involves reference gluons of the same helicity.

\subsec{Outline of Possible Proof}

As stated in the introduction, our conjectured recursion relations
are based on recent results for calculating one-loop $\N=4$
amplitudes \BrittoNC\ and some older results describing their
infrared behavior \irref.  In particular, the infrared behavior is
given by
\eqn\lead{A^{\rm
1-loop}_{n:1}|_{\rm IR} =
\left[ -{1\over \epsilon^2}\sum_{j=1}^n \left(
-t_j^{[2]}\right)^{-\epsilon} \right]A^{\rm
tree}_{n:0}. }
Our ability to  extract from \lead\ a recursion relation from tree
amplitudes is due to the recent discovery in \BrittoNC\ that any
box integral coefficient can be expressed simply as a product of
four tree amplitudes by the formula \eqn\besatwaydi{{\hat
a}_\alpha = {1 \over |{\cal S} |}\sum_{{\cal S},J}  n_J A^{\rm
tree}_1 A^{\rm tree}_2 A^{\rm tree}_3 A^{\rm tree}_4. }
The four amplitudes in the product correspond to the four corners
of the box.  This formula was derived from considering the
quadruple cut in the theory with signature $(++--)$.

{}From the infrared behavior \lead\ it is possible to derive the  relation \sim\ \RoibanIX.
\eqn\sim{ A^{\rm tree}_n = {1\over 2}\sum_{i=1}^{n-3}
B_{1,i+1,n-1,n}.}
Note that each box integral whose coefficient appears in \sim\ has (at least) two adjacent trivalent vertices.

The crucial point of our conjecture is that the factor of $\half$
in \sim\ has a deep meaning. It suggests that the sum splits
naturally into two groups.
 We call them
the ${\cal A}$ and the ${\cal B}$ group. Schematically, one can
formally obtain
\eqn\sesi{ A_n ={1 \over 2} \sum_{i=1}^{n-3}B^{{\cal A}}_{1,i+1,n-1,n}
+ {1 \over 2} \sum_{i=1}^{n-3}B^{\cal B}_{1,i+1,n-1,n}.
}
The two groups are defined and distinguished by the helicity
assignments at the adjacent trivalent vertices. In \besatwaydi,
${\cal S}$ is the set of solutions of momenta in the cut
propagators.  There are two solutions (given explicitly in \BrittoNC), and each determines the
type of helicity assignment ($++-$ or $--+$) allowed at each
trivalent vertex. The ${\cal A}$ group is defined as the one for
which only gluons can circulate in the loop. For the ${\cal B}$ group
the whole $\N=4$ multiplet is allowed (but not necessary
realized).
 (See the appendix for further details.)
This natural separation motivated us to propose the ${\cal A}$ (or
equivalently the ${\cal B}$) conjecture: each set of terms in \sesi\
is enough to reproduce the whole amplitude, i.e.,
\eqn\conje{ A_n = \sum_{i=1}^{n-3}B^{{\cal A}}_{1,i+1,n-1,n} = \sum_{i=1}^{n-3}B^{{\cal B}}_{1,i+1,n-1,n}.}

The ${\cal A}$ part of this conjecture is the recursion relations
proposed in section 2 and used to obtain all the amplitudes in sections 3 and 4. It
is amusing to realize that another (perhaps not very useful)
recursion relation can be written down from the ${\cal B}$ terms. The
tree-amplitudes there would contain fermions and scalars as
external legs. Adding all these terms would give the amplitude
with only gluons.

To complete this proof, it would be necessary to check that the ${\cal A}$ and
${\cal B}$ sums are equal.
 It is natural to expect that this is a consequence of applying Ward
identities to the former to get the latter.  
It would be interesting to pursue this in the future.

It should be possible to derive similar recursion relations for tree amplitudes with external fermions by applying supersymmetry transformations to the recursion relations proposed here.

Since our recursion relation involves tree-level amplitudes, it
should be oblivious to $\N=4$ supersymmetry.  Therefore we believe
that it is most naturally given in terms of  the ${\cal A}$ group.

Equation \besatwaydi\ gives the coefficients $B^{{\cal A}}_{1,i+1,n-1,n}$ in terms of a product
of two three-particle tree amplitudes and two possibly larger tree amplitudes.

The two three-gluon amplitudes can be explicitly reduced to
produce the appropriate conversion factor between scalar box
integrals and scalar box functions.

For details of this derivation, see the appendix.

\subsec{Trivalent-Vertex Representation}

The recursion relation \retrue\ gives a given amplitude in terms
of amplitudes with fewer gluons,
\eqn\retos{\eqalign{ & A_n(1,2,\ldots , (n-1)^-,n^+) = \cr &
\sum_{i=1}^{n-3}\sum_{h=+,-} \left ( A_{i+2}({\hat n},1,2,\ldots
i,-{\hat P}^h_{n,i} ) {1\over P^2_{n,i}} A_{n-i}(+{\hat
P}^{-h}_{n,i}, i+1,\ldots , n-2, {\hat{n-1} } ) \right) .} }
As discussed in section 2, each amplitude on the right hand side
of \retos\ is on-shell and momentum conservation is valid.
Therefore, we can apply the recursion relation again to all of
them to get lower amplitudes. This process can be repeated any
number of times until all amplitudes entering in the expression
for $A_n$ are reduced to three-gluon amplitudes. 
\ifig\hola{Schematic representation of the process to reduce all
tree-level amplitudes to trees with only $(++-)$ and $(--+)$
vertices.} {\epsfxsize=0.95\hsize\epsfbox{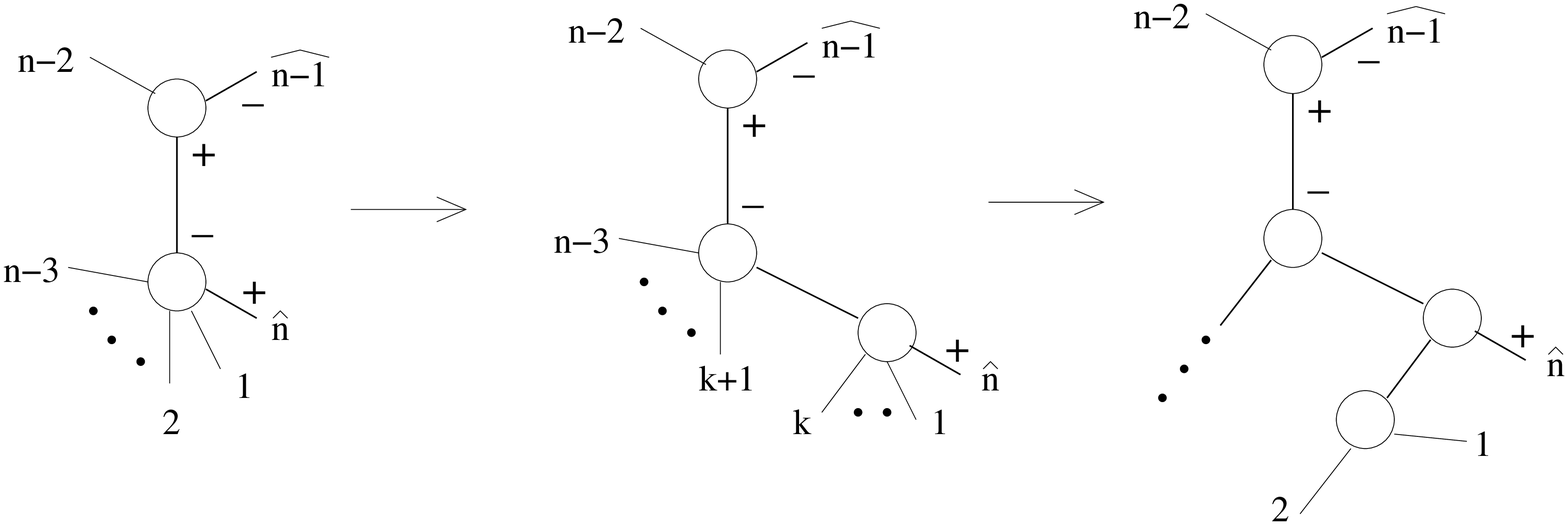}}
 This process is
illustrated in \hola.

It is important to note that this  decomposition is only
possible thanks to the fact that all amplitudes are on-shell and
intermediate momenta take values in signature $(--++)$ where
three-gluon amplitudes on-shell do not vanish.

This decomposition hints at the fascinating possibility that there
might be an effective Lagrangian describing classical gauge theory
in terms of a scalar field with only cubic interactions.

Another tempting conjecture is that the diagrams obtained using
only the amplitudes $(++-)$ and $(--+)$ are the natural outcome of
a string theory whose  target space is the super Calabi-Yau manifold
known as the Quadric. Recall that the original twistor string
theory proposed in \WittenNN\ is a topological B model with target
space the super Calabi-Yau manifold $\Bbb{CP}^{3|4}$. Also in
\WittenNN, an alternative to this target space was suggested. Take
two copies of $\Bbb{CP}^{3|3}$, one with homogeneous coordinates
$Z^I, \psi^A$ and the other with $W^I, \chi^A$ with $I=1,...,4$,
$A=1,...3$. Then the Quadric is the zero  set given by
\eqn\quadric{ \sum_{I=1}^4Z^I W_I + \sum_{A=1}^3 \psi^A\chi_A = 0}
in $\Bbb{CP}^{3|3}\times \Bbb{CP}^{3|3}$.
Note that the original topological B model with target
space  $\Bbb{CP}^{3|4}$ is not manifestly parity invariant and has to be enriched with D-instantons in order to reproduce tree-level amplitudes.  In contrast, the string theory on the Quadric is manifestly parity invariant and requires no D-instantons \WittenNN.

It is also natural to expect that a connection to the MHV diagram
(CSW) construction of tree amplitudes can be made from the
trivalent representation of the amplitudes. The rough idea is
based on two conjectures once the representation in terms of
trivalent vertices is given: 
\ifig\hide{Possible connection between the trivalent
representation of tree amplitudes and MHV diagrams. Note that the
helicities of the gluons connected to each circle have MHV
structure.} {\epsfxsize=0.6\hsize\epsfbox{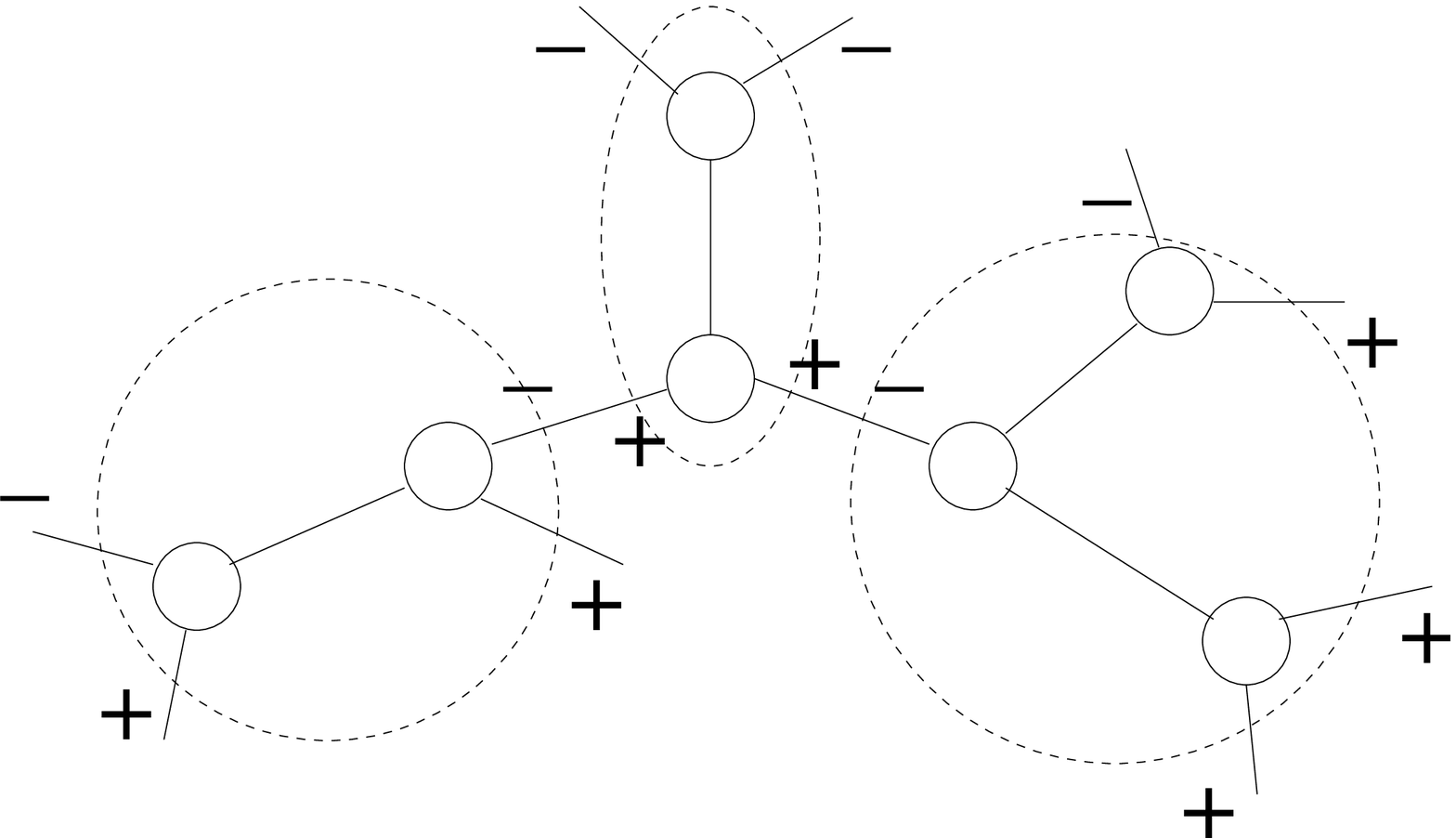}}
One is that it might always be
possible to circle connected set of vertices such that the legs
coming out of the circle have MHV-like helicities (see \hide). 
The second is that all the vertices inside one such circle
will simplify to produce an MHV amplitude. That this might be true
is suggested by the computation in section 3.1, along with the fact that each leg
coming out of the circle is on-shell and the sum of the momenta
is zero.

\subsec{Closed Set of Amplitudes}

The recursion relations \retrue\ give any amplitude in terms of
amplitude with fewer gluons but with generic helicity.
However, it turns out that there is a set of amplitudes that
closes under the recursion procedure. In other words, a given
amplitude in the set is determined only by amplitudes in the set.

The set we found is given by amplitudes of the form
\eqn\colse{ A_{p,q} = A(1^-,2^-,\ldots , p^-, (p+1)^+ , \ldots ,
(p+q)^+)}
for any integers $p\geq 1$ and $q\geq 1$.

Let us study the term contributing to the recursion formula
\retrue\ when the reference gluons are taken to be $p^-$ and
$(p+1)^+$. It is not difficult to check that only two terms are
nonzero. They are given by
\eqn\seto{\eqalign{ & ((p-1), \hat{p}~|~ \widehat{p+1},\ldots
,(p+q),1,\ldots , (p-2)), \cr & ((p+3),\ldots , (p+q),
1,\ldots ,\hat{p}~|~\widehat{p+1},(p+2)). } }

The first term is given by $A_{p-1,q}$ times $A_{2,1}$. The second
term is given by $A_{p,q-1}$ times $A_{1,2}$.

This proves that the set of amplitudes \colse\ closes under
\retrue. It would be very interesting to find an explicit solution
to these equations.

One simple observation is that the number of terms in $A_{p,q}$,
which we denote by $N_{p,q}$, satisfy the following recursion
relation: $ N_{p,q}= N_{p-1,q}+N_{p,q-1} $ with boundary
conditions $N_{2,q}=1,~\forall q\geq 1$ and $N_{p,2}=1,~\forall
p\geq 1$. 
Thus we recognize the number of terms as a binomial coefficient:
$N_{p,q} = (p+q-4)! / ((p-2)!(q-2)!)$.

A special case of this closed set is the next-to-MHV amplitudes in 
the helicity configuration $(---++ \cdots +)$.  These amplitudes have been written down in \refs{\KosowerXY,\CachazoKJ,\KosowerYZ}.
It would be interesting to check that our recursion relations reproduce these existing results.

\subsec{ New Recursion Relations and Linear Trees}

  Finally, we want to mention another interesting direction. Recall
that the recursion relations \retrue\ use as reference vectors two
gluons of opposite helicity. One natural question to ask is
whether the same formula is valid for reference gluons of the same
helicity. We have evidence that this is indeed the case, we have
computed all next-to-MHV six-gluon amplitudes (except, of course,
$A(1^+,2^-,3^+,4^-,5^+,6^-)$) and found perfect agreement.

In section 5.2, we described how to iterate \retrue\ in order to
write any tree amplitude in terms of trivalent vertices. The final
answer is given in terms of general trees with trivalent vertices
(see \hola). But now we can do better. We can use any two
gluons as reference vectors, regardless of their helicity, and make
the following sequence of choices, $((n-1),n)$, $(n,1)$, $(1,2)$, and so
on. By doing this we produce only linear trees! It would be
interesting to prove that this simple picture reproduces all known
amplitudes.

\bigskip
\bigskip
\centerline{\bf Acknowledgments}

It is a pleasure to thank M. Spradlin, P. Svr\v{c}ek, A. Volovich and E. Witten for useful
discussions. R. B. and B. F. were supported by NSF grant
PHY-0070928. F. C. was supported in part by the Martin A. and
Helen Chooljian Membership at the Institute for Advanced Study and
by DOE grant DE-FG02-90ER40542.

\appendix{A}{From box integrals to tree amplitudes}

Here we show how the coefficient $B_{1,i+1,n-1,n}$ may be written as the product of two tree amplitudes times a propagator.  We follow the notation and conventions of \BrittoNC, which includes 
the necessary background material.
 In particular, we will use the shorter notation $d_{r,i}$ to represent these coefficients, where in contrast to our previous notation, $r$ may now take values all the way from $1$ to $n-3$.
In this appendix, we fix the  massless legs to be $(i-2)$ and $(i-1)$.

The plan of this appendix is the following. First we solve the
loop momenta $\ell_i$, $i=1,2,3,4$ explicitly. Then we
find the box coefficients for the configuration where
both  massless legs $(i-2)$ and $(i-1)$ have equal (positive) helicities.
Finally we find the box coefficients for the configuration
$(i-2)^-~(i-1)^+$ in terms of two tree amplitudes times a propagator.

\subsec{Solving for the loop momentum}

Here we solve for the loop momentum explicitly. Let us
start from the configuration (a), where  the $K_3$ vertex has
helicity distribution $(++-)$ while the $K_4$ vertex has helicity
distribution $(--+)$.  We must then have the following relationships (for
definitions of $\ell_i,K_i$, see Figure 5):
\eqn\aloopone{\eqalign{& \lambda_{\ell_4}  =  \a \lambda_{i-2},~~~~~
\lambda_{\ell_3}  =  \b \lambda_{i-2},~~~~~\cr
& \tilde\lambda_{\ell_4}  =  \gamma \tilde\lambda_{i-1},~~~~~
\tilde\lambda_{\ell_1}  =  \rho \tilde\lambda_{i-1},\cr
& \b \tilde\lambda_{\ell_3}  =  \tilde \lambda_{i-2}+\a\gamma
\tilde\lambda_{i-1},
~~~~~
\rho \lambda_{\ell_1}  = \a\gamma  \lambda_{i-2}-\lambda_{i-1}.}}
Using $\ell_2^2=0$, we find that
\eqn\alooptwo{\a \gamma= { K_{23}^2\over \gb{i-2|K_2|i-1}}=
{ K_{14}^2\over -\gb{i-2|K_1|i-1} }.  }
To make the calculation easier, we can use scaling freedom to
fix $\a=\b=\rho=1$, so that we have solved the loop momenta as
the following:
\eqn\aloopthree{\eqalign{& \lambda_{\ell_4}  =  \lambda_{i-2},
~~~~~\tilde\lambda_{\ell_4}=\gamma\tilde\lambda_{i-1},~~~~~~
\ell_4= \gamma\lambda_{i-2}\tilde\lambda_{i-1}, \cr
& \lambda_{\ell_3}  =  \lambda_{i-2},~~~~~\tilde\lambda_{\ell_3}=
\tilde \lambda_{i-2}+\gamma
\tilde\lambda_{i-1},~~~~~~\ell_3=p_{i-2}+\ell_4, \cr
& \lambda_{\ell_1}  =  \gamma  \lambda_{i-2}-\lambda_{i-1},~~~~~
\tilde\lambda_{\ell_1}  =   \tilde\lambda_{i-1},~~~~~~\ell_1=
\ell_4-p_{i-1}, \cr
& \gamma  =  { K_{23}^2\over \gb{i-2|K_2|i-1}}=
{ K_{14}^2\over -\gb{i-2|K_1|i-1}}.   }}
In particular, we find that
\eqn\aloopfour{\ell_2 = \ell_4-K_{14}= -\left( K_{14}-
{K_{14}^2\over 2 q\cdot K_{14}} q\right),}
where $q$ is a new momentum defined by
$q_{\a\dot{\a}}=(\lambda_{i-2})_{\a}(\tilde \lambda_{i-1})_{\dot{\a}}$.
The meaning inside the parentheses of \aloopfour\ is simply the projection
of the massive momentum $K_{14}$ along the direction defined by our two reference spinors.
If we call the $\ell_4={K_{14}^2\over 2 q\cdot K_{14}} q$
the transformation of momentum, we see that from \aloopthree\
$\ell_1$ is the momentum $-p_{i-1}$ transformed by $\ell_4$. Similarly,
$\ell_3$ is a transformed version of $p_{i-2}$.

Now we move to the configuration (b) where the $K_3$ vertex has
helicity distribution $(--+)$ while the $K_4$ vertex has
helicity distribution $(++-)$. Using a similar method, we get the
following result:
\eqn\bloopone{\eqalign{&
\lambda_{\ell_4}  = \lambda_{i-1},~~~~~\tilde\lambda_{\ell_4}
=\a\tilde\lambda_{i-2},~~~~~\ell_4=\a  \lambda_{i-1}
\tilde\lambda_{i-2}, \cr
& \lambda_{\ell_3}  =  \lambda_{i-2}+\a \lambda_{i-1},~~~~~~
\tilde\lambda_{\ell_3}  =  \tilde \lambda_{i-2},
~~~~~\ell_3=p_{i-2}+\ell_4, \cr
& \lambda_{\ell_1}  =  \lambda_{i-1},~~~~~\tilde\lambda_{\ell_1}  =
\a  \tilde\lambda_{i-2}- \tilde\lambda_{i-1},
~~~~~\ell_1=\ell_4-p_{i-1}, \cr
& \a   =  { K_{23}^2\over \gb{i-1| K_2|i-2}}=
 -{ K_{14}^2\over \gb{i-1| K_1|i-2}}, }}
and
\eqn\blooptwo{ \ell_2 = \ell_4-K_{14}= -\left( K_{14}-
{K_{14}^2\over 2 \tilde q\cdot K_{14}} \tilde q\right),}
where $\tilde q$ is a new momentum defined by
$\tilde q_{\a\dot{\a}}=(\lambda_{i-1})_{\a}(\tilde
\lambda_{i-2})_{\dot{\a}}$ so that $\ell_2$ is a light-cone projection the massive
momentum $K_{14}$.
The momenta $q$ and $\tilde q$ are related by
conjugation. 

\subsec{The box coefficients: I}

In this part we find  the coefficients of two-mass-hard box functions
 where both  massless legs $(i-2)$ and $(i-1)$ have equal
(positive) helicities.

\ifig\kkobp{The two possible helicity configurations of
 box functions
 where both  massless legs $(i-2)$ and $(i-1)$ have equal
positive helicities.
}
{\epsfxsize=0.70\hsize\epsfbox{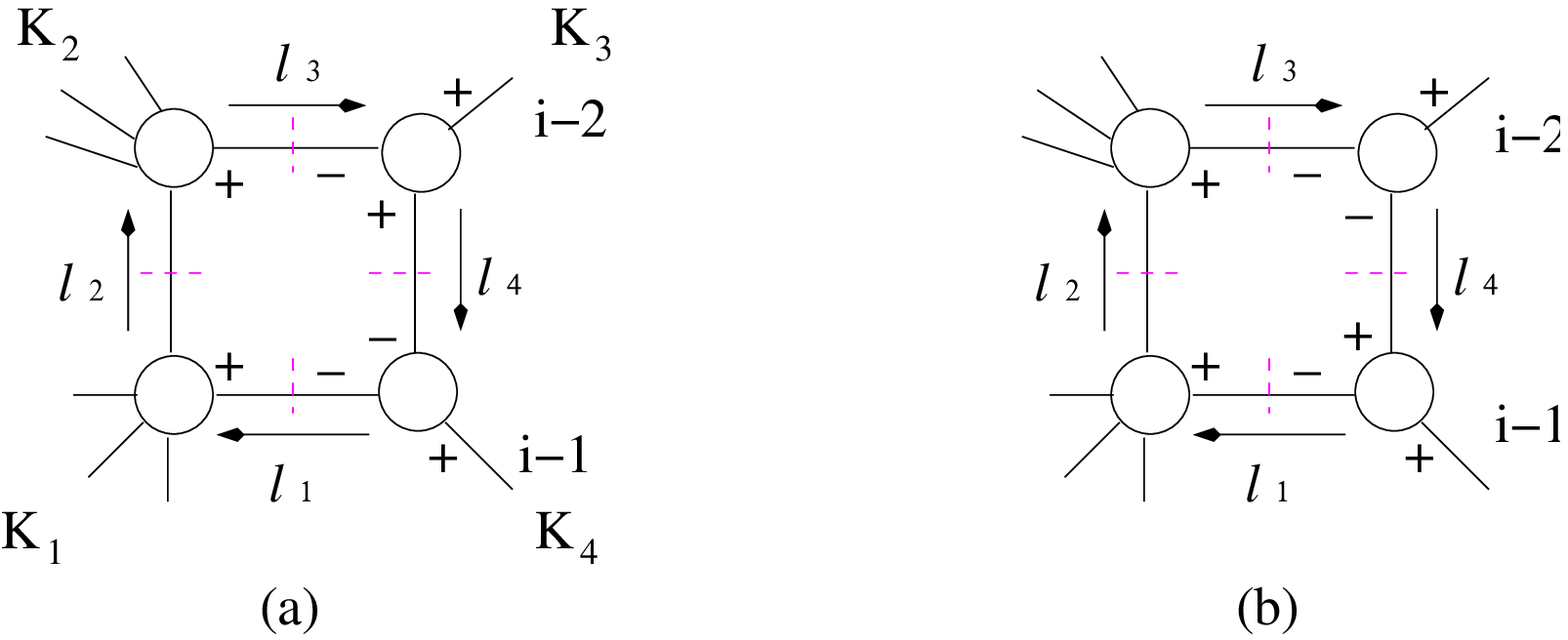}}

With both  massless legs $(i-2)$ and $(i-1)$ having positive helicities, there are two possible
contributions with different helicity assignments
(see \kkobp). Since we have solved the loop
momenta $\ell_1,\ell_2,\ell_3,\ell_4$ in the previous subsection,
 we can use the
general formula from the quadruple cut to read off the
corresponding coefficients.

As given in \BrittoNC, the coefficient of the box integral $I^{2m~h}$
is calculated by
\eqn\besatwaydiy{{\hat d}_\alpha = {1 \over |{\cal S}|}\sum_{\cal S} A^{\rm
tree}_1 A^{\rm tree}_2 A^{\rm tree}_3 A^{\rm tree}_4.
}
The coefficient of $F^{2m~h}$
is then given by $d_\alpha=-2{\hat d}_\alpha/(K_{34}^2 K_{14}^2)$.

Consider first the configuration (a) of  \kkobp.
Since in this situation
 only gluons may circulate in the loop, we find that
$$
{\hat d}_a  =  { A_{\rm tree}(K_4) A_{\rm tree}(K_1) A_{\rm tree}(K_2)
A_{\rm tree}(K_3) \over 2}.
$$
We use the solution \aloopthree\ to determine that 
$$
A_{\rm tree}(K_3)=\gamma [i-2~i-1], \quad 
A_{\rm tree}(K_4)=-{\vev{i-2~i-1}\over \gamma}.
$$
Since $K_{34}^2=\vev{i-2~i-1}[i-2~i-1]$, we find the 
result
\eqn\acoeff{d_a =
{ A_{\rm tree}(-\ell_1^+,K_1,+\ell_2^\pm) A_{\rm tree}(-\ell_2^{\mp},K_2,
+\ell_3^+) \over K_{14}^2}, }
where the loop momenta are given by \aloopthree\ and
\aloopfour, i.e.
\eqn\acoeffmom{ -\ell_1  =   p_{i-1}-\delta p,~~~~\ell_3=p_{i-2}+\delta p,
~~~~~\ell_2= -K_{14}+\delta p,~~~~\delta p=
{K_{14}^2\over 2 q\cdot K_{14}} q}
with
$q_{\a\dot{\a}}=(\lambda_{i-2})_{\a}(\tilde \lambda_{i-1})_{\dot{\a}}$.

For configuration (b)  of  \kkobp\ we do a similar calculation and get the
coefficient
\eqn\bcoeff{d_b =
{ A_{\rm tree}(-\ell_1^+,K_1,+\ell_2^\pm) A_{\rm tree}(-\ell_2^{\mp},K_2,
+\ell_3^+)\over  K_{14}^2}}
where the loop momenta are given by \bloopone\ and \blooptwo, i.e.,
\eqn\bcoeffmom{-\ell_1  =   p_{i-1}-\delta \tilde p,~~~~~
\ell_3=p_{i-2}+\delta  \tilde p,~~~~~
\ell_2= -K_{14}+\delta  \tilde p,~~~~\delta  \tilde p =
{K_{14}^2\over 2 \tilde q\cdot K_{14}} \tilde q}
with $\tilde q_{\a\dot{\a}}=(\lambda_{i-1})_{\a}(\tilde
\lambda_{i-2})_{\dot{\a}}$.

The overall coefficient is the sum of the
two contributions in \acoeff\ and \bcoeff.

\subsec{The box coefficients: II}

\ifig\kkobptwo{Three possible helicity configurations
where part (a) can only have gluons circulating in the 
loop while parts (b) and (c) can have fermions and
complex scalars circulating.
}
{\epsfxsize=0.85\hsize\epsfbox{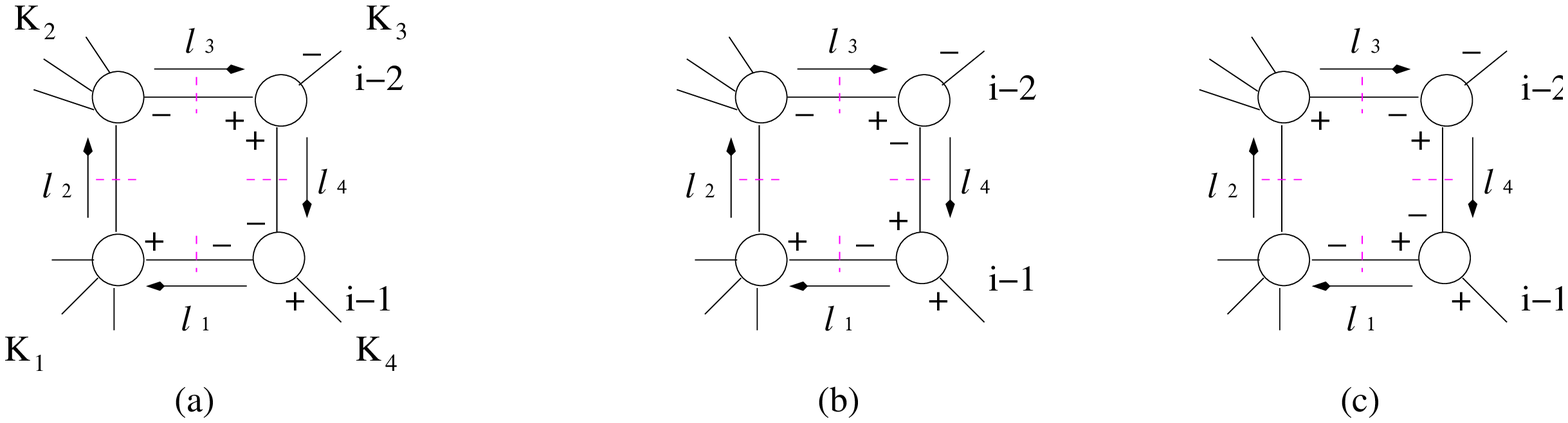}}

In this subsection, we present the  coefficients of two-mass-hard box functions
 where   massless leg $(i-2)$ has negative helicity
and massless leg $(i-1)$ has
positive helicity. For this assignment of helicities there are
three possible configurations (see \kkobptwo).
The calculations are similar to the previous case, so we will be brief and mention only the new features.

For configuration (a), only gluons are allowed to circulate in
the loop. With the exact same calculation we reach
\eqn\MPacoeff{d_a =
{ A_{\rm tree}(-\ell_1^+,K_1,+\ell_2^\pm) A_{\rm tree}(-\ell_2^{\mp},K_2,
+\ell_3^-) \over K_{14}^2}, }
where the loop momenta are given by \aloopthree\ and
\aloopfour.

For configuration (b), depending on the helicity distribution
at the vertices $K_1$ and $K_2$,
it is possible that chiral fermions and
complex scalars of ${\cal N}=4$ vector multiplet circulate in 
the loop. Thus we get the contribution of
part (b) with helicity solution given by \bloopone\ and \blooptwo\ as
\eqn\MPbcoeff{d_b  =  \sum_{a=0,1,2} n_a
 { A_{tree}^a(-\ell_1^+,K_1,+\ell_2^{\pm}) A_{tree}^a
(-\ell_2^{\mp},K_2,+\ell_3^-)\over  K_{14}^2} \a^a,}
where $a=0,1,2$ are for gluons, fermions and scalars and $n_a$
are $1$, $-4$ and $3$ respectively. For example,
when $a=1$, the particles propagating with momenta $\ell_1$, $\ell_2$ and $\ell_3$ are all fermions.

Configuration (c) is similar to configuration (b).
With the helicity solution given by \bloopone\ and \blooptwo\ we
find the contribution to coefficients from part (c) to be
\eqn\MPccoeff{d_c  =  \sum_{a=0,1,2} n_a
{ A_{tree}^a(-\ell_1^-,K_1,+\ell_2^{\pm}) A_{tree}^a
(-\ell_2^{\mp},K_2,+\ell_3^+)\over  K_{14}^2} \a^{4-a}.}

The overall coefficient is given by the sum of
\MPacoeff, \MPbcoeff\ and \MPccoeff.

As mentioned in the main text, it is the formula \MPacoeff\
that inspired us to make the conjecture \retrue. The relationship
between the two formulas \MPacoeff\ and \retrue\ is the 
following: $-\ell_1\to \widehat{n}$, $\ell_2\to -\widehat{P}_{n,i}$
and $\ell_3 \to \widehat{n-1}$.
In principle, we can use formula \acoeff\ or \bcoeff\ to
calculate the tree level amplitude if we take the reference gluons to be
$(i-2)^+$ and $(i-1)^+$. The reader can also find the expressions
for coefficients of box functions with helicities
$(i-2)^-~(i-1)^-$ or $(i-2)^+~(i-1)^-$, which are related to
the above two solutions by conjugation.


\listrefs

\end